\documentclass{article}

\usepackage[utf8]{inputenc}
\usepackage[T1]{fontenc}

\RequirePackage{amsthm,amsmath,amsfonts,amssymb}
\usepackage{bm}
\RequirePackage[authoryear]{natbib}
\usepackage{multirow}
\RequirePackage[colorlinks,citecolor=blue,urlcolor=blue,backref=page,backref=page]{hyperref}
\RequirePackage{graphicx}
\usepackage{algorithm}
\usepackage{algpseudocode}
\usepackage{booktabs}

\algnewcommand\algorithmicinput{\textbf{Input:}}
\algnewcommand\Input{\item[\algorithmicinput]}
\algnewcommand\algorithmicoutput{\textbf{Output:}}
\algnewcommand\Output{\item[\algorithmicoutput]}
\algnewcommand{\IfThenElse}[3]{
  \State \algorithmicif\ #1\ \algorithmicthen\ #2\ \algorithmicelse\ #3}

\algdef{SE}[SUBALG]{Indent}{EndIndent}{}{\algorithmicend\ }%
\algtext*{Indent}
\algtext*{EndIndent}

\title{Sequential Rank and Preference Learning with the Bayesian Mallows Model}
\author{
Øystein Sørensen\thanks{Department of Psychology, University of Oslo. \texttt{oystein.sorensen@psykologi.uio.no}} \and
Anja Stein\thanks{School of Mathematical Sciences, Lancaster University. \texttt{a.k.stein@outlook.com}} \and
Waldir Leoncio Netto\thanks{Oslo Centre for Biostatistics and Epidemiology, University of Oslo. \texttt{w.l.netto@medisin.uio.no}} \and
David S. Leslie\footnotemark[2] 
}
\date{}

\begin{document}
\maketitle

\begin{abstract}
The Bayesian Mallows model is a flexible tool for analyzing data in the form of complete or partial rankings, and transitive or intransitive pairwise preferences. In many potential applications of preference learning, data arrive sequentially and it is of practical interest to update posterior beliefs and predictions efficiently, based on the currently available data. Despite this, most algorithms proposed so far have focused on batch inference. In this paper we present an algorithm for sequentially estimating the posterior distributions of the Bayesian Mallows model using nested sequential Monte Carlo. The algorithm requires minimal user input in the form of tuning parameters, is straightforward to parallelize, and returns the marginal likelihood as a direct byproduct of estimation. We evaluate its performance in simulation experiments, and illustrate a real use case with sequential ranking of Formula 1 drivers throughout three seasons of races.
\end{abstract}

\section{Introduction}

Data in the form of rankings and preferences arise naturally across a variety of domains. Examples include content recommendation based on click data \citep{liuDiversePersonalizedRecommendations2019}, algorithm comparison \citep{rojas-delgadoBayesianPerformanceAnalysis2022}, consumer preferences \citep{courcouxPreferenceDataAnalysis1997,kamishimaNantonacCollaborativeFiltering2003,krivulinUsingPairwiseComparisons2022,manuelConsumersPerceptionsPreference2015}, grant panel reviews \citep{pearceUnifiedStatisticalLearning2022}, genome-wide transcriptomic analyses \citep{eliseussenRankbasedBayesianVariable2022,vitelliTranscriptomicPancancerAnalysis2023}, analysis of mutual funds' preferences for governance structures \citep{yiWhichFirmsRequire2021}, social hierarchies \citep{nichollsBayesianInferencePartial2022}, and reinforcement learning from human feedback \citep{hwangSequentialPreferenceRanking2023}. An early model for analyzing rankings \citep{thurstoneLawComparativeJudgment1927} assumed a judge ranks $m$ items by assigning a score to each item, and then ordering them according to the score. Further developments of this approach include the Plackett-Luce model \citep{luceIndividualChoiceBehavior1959,plackettAnalysisPermutations1975}, the Babington Smith model \citep{babingtonsmithDiscussionProfessorRoss1950}, and the Bradley-Terry model \citep{bradleyRankAnalysisIncomplete1952}, all of which are based on assigning real-valued utilities to each item, yielding a large number of parameters to be estimated.

Now consider a collection of items $\mathcal{A} = \{A_{1}, A_{2}, \dots, A_{m}\}$, and let $\bm{\rho}$ be a permutation of the integers $[m]:=\{1,2,\dots,m\}$ denoting the items' modal ranking in the population of interest, such that $\rho_{i}$ denotes the modal ranking of  item $A_{i}$. For a particular individual, let $A_{i} \succ A_{j}$ imply that the individual prefers $A_{i}$ to $A_{j}$, and let $\bm{r}$ be the permutation of $[m]$ encoding the individual's rankings. \citet{mallowsNonNullRankingModels1957} showed that if the probability that an individual ranks a pair of items in agreement with their relative position in $\bm{\rho}$ is given by 
\begin{equation*}
    P\left(A_{i} \succ A_{j} | \rho_{i} < \rho_{j}\right) = 0.5 + 0.5 ~\text{tanh}\left\{(\rho_{j} - \rho_{i}) \log \theta + \log \phi \right\}
\end{equation*}
we obtain an exponential model $P(\bm{r} | \bm{\rho}) \propto \exp\{-\alpha d(\bm{\rho}, \bm{r})\}$ for the observed ranking $\bm{r}$. When $\phi=1$, $d(\cdot, \cdot)$ is Spearman's rank correlation \citep{spearmanProofMeasurementAssociation1904} and when $\theta=1$, $d(\cdot, \cdot)$ is the Kendall distance \citep{kendallNewMeasureRank1938}. The precision parameter $\alpha$ quantifies how far observed rankings typically are from the modal ranking. Advantages of the Mallows model over utility-based models include a lower number of parameters ($\alpha$ and $\bm{\rho}$) and the fact that its support is defined on the space of rankings. The model has later been extended to incorporate additional distance functions \citep{diaconisGroupRepresentationsProbability1988} and to item-dependent precision parameters \citep{flignerDistanceBasedRanking1986}.  We refer to the reviews by \citet{liuModelBasedLearningPreference2019} and \citet{yuAnalysisRankingData2019} and the monograph by \citet{mardenAnalyzingModelingRank1995} for further details.

\citet{vitelliProbabilisticPreferenceLearning2017} proposed a Bayesian Mallows model and a Markov chain Monte Carlo (MCMC) algorithm for its estimation. Compared to other approaches focusing on the Kendall or Cayley distances  \citep{irurozkiSamplingLearningMallows2018,luEffectiveSamplingLearning2014,meilaExponentialModelInfinite2010}, \citet{vitelliProbabilisticPreferenceLearning2017}'s algorithm works naturally with any of the distance functions proposed by \citet{diaconisGroupRepresentationsProbability1988} for the Mallows model, and it incorporates data in the form of partial rankings or pairwise preferences. Its fully Bayesian approach allows predicting users' preferences of items they have not yet seen, allowing the model to be used as a probabilistic recommender system \citep{liuDiversePersonalizedRecommendations2019}.

The MCMC algorithm of \citet{vitelliProbabilisticPreferenceLearning2017} has some drawbacks, however. The user has to set tuning parameters for the proposal distributions for precision parameters, modal rankings, and latent rankings. It hence may require several pilot runs for obtaining sufficient acceptance probabilities. Second, in settings where data arrive sequentially, updating the posteriors requires running the full algorithm from scratch. The goal of this paper is to alleviate these issues. To this end, we propose a nested sequential Monte Carlo (SMC$^{2}$) algorithm \citep{chopinSMC2EfficientAlgorithm2013,fulopEfficientLearningSimulation2013}, which requires minimum user input and efficiently updates posterior distributions when new data arrive. The algorithm is straightforward to parallelize and is an extension of the work by  \citet{steinSequentialInferenceMallows2023}, who proposed a sequential Monte Carlo (SMC) algorithm using a resample-move scheme \citep{berzuiniRESAMPLEMOVEFilteringCrossModel2001,chopinSequentialParticleFilter2002,gilksFollowingMovingTarget2001}.

\begin{figure}
    \centering
    \includegraphics[width=\linewidth]{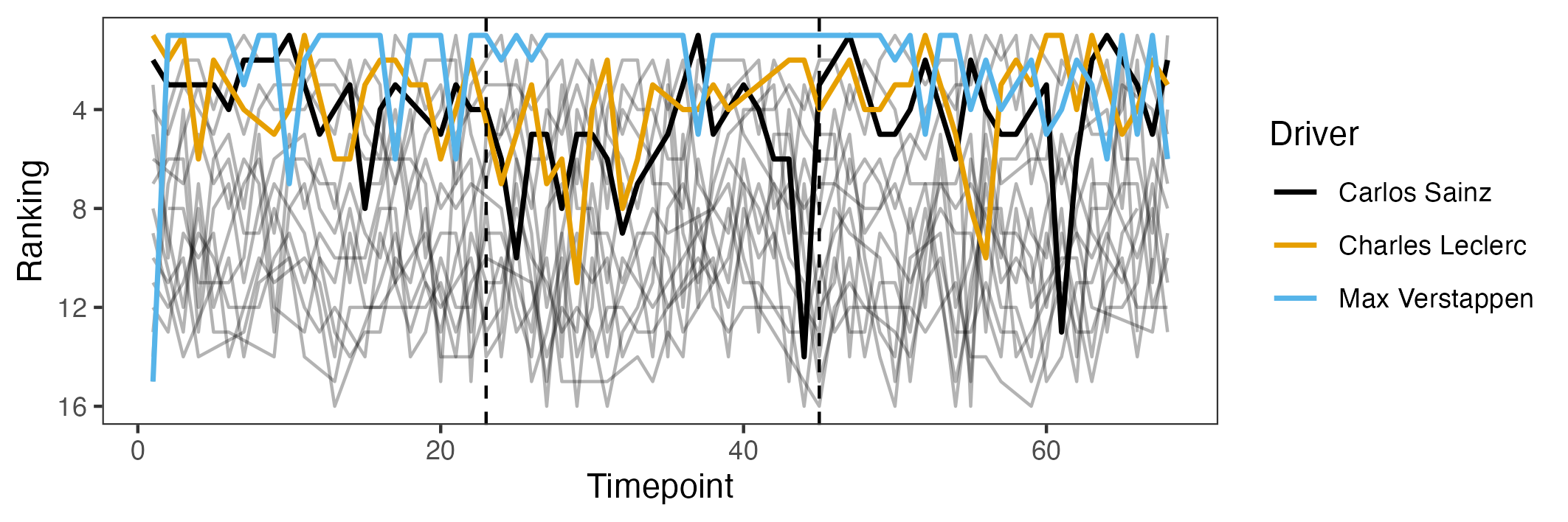}
    \caption{Rankings of drivers based on race results in Formula 1 seasons 2022, 2023, and 2024. Vertical dashed lines indicate the start of a season. Traces for the three drivers with best mean ranking across all races are shown in color.}
    \label{fig:f1-data-plot}
\end{figure}

A motivating example is shown in Figure \ref{fig:f1-data-plot}, showing race results from the Formula 1 seasons 2022-2024.\footnote{Data were downloaded from \href{https://github.com/toUpperCase78/formula1-datasets}{https://github.com/toUpperCase78/formula1-datasets}.} We included the 16 drivers who completed 50 or more of the total 68 races, and computed rankings from the results of each race. If we assume that the underlying skills of the drivers -- as well as the performance of the cars and teams -- are relatively stable across the three seasons, the results of each race can be viewed as a noisy assessment of the underlying true ranking. Using the SMC$^{2}$ algorithm proposed in this paper we can easily update the posterior distribution for the ranking of the drivers after each race. We will revisit these data in Section \ref{sec:applications}.

The paper proceeds as follows. In Section \ref{sec:Background} we provide necessary background on the Bayesian Mallows model. In Section \ref{sec:SequentialInference} we propose an SMC$^{2}$ algorithm for the Bayesian Mallows model with partial rankings or pairwise preference data. In Section \ref{sec:TopologicalOrderings} we investigate the computational requirements for computing topological orderings -- an essential part of the algorithms. In Section \ref{sec:SimulationExperiments} we report the results of simulation experiments and in Section \ref{sec:applications} we revisit the Formula 1 data. We conclude and discuss further developments in Section \ref{sec:Discussion}. An R  \citep{rcoreteamLanguageEnvironmentStatistical2024} package providing an API to our C++ implementation is available from GitHub\footnote{\href{https://github.com/osorensen/BayesMallowsSMC2}{https://github.com/osorensen/BayesMallowsSMC2}} and R code for reproducing all the results in the paper is available from our OSF repository.\footnote{\href{https://osf.io/pquk4/}{https://osf.io/pquk4/}}

\section{Background and Model Setup}
\label{sec:Background}

We now introduce the Bayesian Mallows model as it was defined in \citet{vitelliProbabilisticPreferenceLearning2017} and \citet{crispinoBayesianMallowsApproach2019}. As the goal of this paper is to develop generic algorithms we present the model in full generality, and do not discuss modeling choices. Any particular application will typically use special cases of the presented framework.

\subsection{Mallows' Model for Partial Rankings and Pairwise Preferences}
\label{sec:MallowsBackground}
Let $\mathcal{P}_{m}$ denote the space of all permutations of $[m]$ and consider rankings $\bm{r} \in \mathcal{P}_{m}$ of a set of items $\mathcal{A}$ distributed according to a mixture of Mallows models \citep{diaconisGroupRepresentationsProbability1988,mallowsNonNullRankingModels1957,vitelliProbabilisticPreferenceLearning2017} with $C$ components
\begin{equation}
\label{eq:MallowsMixture}
p\left(\bm{r} | \theta\right) = \sum_{c=1}^{C} \tau_{c} Z\left(\alpha_{c}\right)^{-1} \exp\left\{-\alpha_{c}  d\left(\bm{r}, \bm{\rho}_{c}\right)\right\} 1\left\{ \bm{r} \in \mathcal{P}_{m} \right\},
\end{equation}
where $\theta = \{\alpha_{c}, \bm{\rho}_{c}, \tau_{c}\}_{c=1}^{C}$ and $1\{A\}$ is an indicator function for the event $A$. For each cluster $c$, $\alpha_{c} \in \mathbb{R}_{\geq 0}$ is a precision parameter and $\bm{\rho}_{c} \in \mathcal{P}_{m}$ is the modal ranking. $d(\cdot, \cdot)$ is a right-invariant distance function, and 
\begin{equation}
    \label{eq:MallowsNormalizingConstant}
    Z\left(\alpha\right) = \sum_{\bm{r} \in \mathcal{P}_{m}} \exp\left\{-\alpha d\left(\bm{r}, \bm{e}\right) \right\}
\end{equation}
is the normalizing constant where $\bm{e} = (1, 2, \dots, m)'$. Subject to the constraint $\sum_{c=1}^{C} \tau_{c} = 1$, $\tau_{c} \in [0,1]$ is the proportion of the population belonging to the $c$'th cluster.

A ranking $\bm{r}$ is a latent variable, and observations $\bm{y}$ are distributed according to $p(\bm{y} | \bm{r}, \theta)$. The marginal likelihood of $N$ observations $\bm{y}_{1:N} = \{\bm{y}_{1}, \bm{y}_{2}, \dots, \bm{y}_{N} \}$ is
\begin{equation}
\label{eq:MarginalLikelihood}
    p\left(\bm{y}_{1:N} | \theta\right) = \prod_{n=1}^{N} \sum_{c=1}^{C} \tau_{c} Z\left(\alpha_{c}\right)^{-1} \sum_{\bm{r}_{n} \in \mathcal{P}_{m}}
     \exp\left\{-\alpha_{c} d\left(\bm{r}_{n}, \bm{\rho}_{c}\right)\right\} p_{\epsilon}\left(\bm{y}_{n} |\bm{r}_{n}\right)
\end{equation}
where $p_{\epsilon}(\bm{y}_{n} | \bm{r}_{n})$ is the sampling distribution of the observed rankings given the latent rankings where $\epsilon$ is an error parameter to be introduced later. Here and in the sequel, Greek letter subscripts imply conditioning. 

Complete rankings correspond to $p_{\epsilon}(\bm{y}_{n} | \bm{r}_{n}) = 1\{\bm{y}_{n} = \bm{r}_{n}\}$. For top-$k_{n}$ rankings with $k_{n} \in [m]$ we define the set of items ranked by user $n$ as $\mathcal{A}_{n} = \{A_{i} \in \mathcal{A} : r_{ni} \leq k_{n}\}$ whereas for ranks missing completely at random we let $\mathcal{A}_{n}$ define the set of ranked items. In both cases we have $p_{\epsilon}(y_{ni} | \bm{r}_{n}) = 1\{y_{ni} = r_{ni}\}$ for $i : A_{i} \in \mathcal{A}_{n}$. Defining the set of latent rankings consistent with the observations
\begin{equation}
\label{eq:ConstraintSetPartial}
    \mathcal{S}_{n} = \left\{ \bm{r} \in \mathcal{P}_{m} : \left( r_{i} = y_{ni}~ \forall  i :  A_{i} \in \mathcal{A}_{n}  \right)\right\}
\end{equation}
with the complete data case given by $\mathcal{S}_{n} = \{\bm{y}_{n}\}$, the marginal likelihood \eqref{eq:MarginalLikelihood} reduces to
\begin{equation}
\label{eq:MarginalLikelihoodMissing}
    p\left(\bm{y}_{1:N} | \theta\right) = \prod_{n=1}^{N} \sum_{c=1}^{C} \tau_{c} Z\left(\alpha_{c}\right)^{-1} \sum_{\bm{r}_{n} \in \mathcal{S}_{n}}
     \exp\left\{-\alpha_{c} d\left(\bm{r}_{n}, \bm{\rho}_{c}\right)\right\}.
\end{equation}

When the data contain pairwise preferences for $p_{n} \leq \binom{m}{2}$ pairs of items, let $y_{ni}$ denote the $i$'th pairwise preference of user $n$. We define the function
\begin{equation*}
    g\left(y_{ni}, \bm{r}_{n}\right) =
    \begin{cases}
        0 & \text{if } y_{ni} = \left(A_{s} \succ A_{t}\right) \text{ and } \left(r_{ns} < r_{nt}\right) \\
     1 & \text{if } y_{ni} = \left(A_{s} \succ A_{t}\right) \text{ and } \left(r_{ns} > r_{nt}\right)
    \end{cases}
\end{equation*}
indicating whether the pairwise preferences contradict the latent rankings. Assuming a true latent ranking exists, inconsistencies arise due to errors made by the users.\footnote{This assumption is indeed a mathematical idealization, as empirical evidence suggest that human preferences are inherently non-transitive \citep{kahnemanProspectTheoryAnalysis1979}.} If errors occur independently at rate $\epsilon \in [0, 1]$, the direction of the preference relation for the $i$'th pair in $\bm{y}_{n}$ has distribution
\begin{equation*}
 p_{\epsilon}\left(y_{ni} | \bm{r}_{n}\right) =
 \begin{cases}
     1 - \epsilon & \text{if } g\left(y_{ni}, \bm{r}_{n}\right) = 0 \\
     \epsilon & \text{if } g\left(y_{ni}, \bm{r}_{n}\right) = 1.
 \end{cases}
\end{equation*}
Setting $\epsilon > 0$ allows mutually incompatible preferences \citep{crispinoBayesianMallowsApproach2019}, and the marginal likelihood \eqref{eq:MarginalLikelihood} becomes
\begin{align*}
    & p\left(\bm{y}_{1:N} | \theta\right) = \\
    & \prod_{n=1}^{N} \sum_{c=1}^{C} \tau_{c} Z\left(\alpha_{c}\right)^{-1} \sum_{\bm{r}_{n} \in \mathcal{P}_{m}}
     \exp\left\{-\alpha_{c} d\left(\bm{r}_{n}, \bm{\rho}_{c}\right)\right\} \left(\frac{\epsilon}{1 - \epsilon}\right)^{\sum_{i=1}^{p_{n}} g(y_{ni}, \bm{r}_{n})} \left(1 -\epsilon\right)^{p_{n}}
\end{align*}
where $\theta$ now also contains $\epsilon$. To only allow mutually compatible pairwise preferences we set $\epsilon = 0$ and define the set of latent rankings consistent with $\bm{y}_{n}$ as 
\begin{equation}
\label{eq:ConstraintSetPairwise}
    \mathcal{S}_{n} = \left\{ \bm{r} \in \mathcal{P}_{m} : \left(A_{s} \succ A_{t} \right) \in \text{tc}\left(\bm{y}_{n}\right) \Leftrightarrow r_{s} < r_{t} \right\},
\end{equation}
where $\text{tc}(\bm{y}_{n})$ is the transitive closure of the directed acyclic graph induced by the pairwise preferences. In this case we can compute the marginal likelihood using \eqref{eq:MarginalLikelihoodMissing}, replacing $\mathcal{S}_{n}$ from \eqref{eq:ConstraintSetPartial} with $\mathcal{S}_{n}$ from \eqref{eq:ConstraintSetPairwise}.

\subsection{Prior Distributions}
\label{sec:BayesianMallowsModel}

For the precision parameters $\alpha_{c}$ we follow \citet{crispinoBayesianMallowsApproach2019} and use independent gamma priors with shape $\gamma > 0$ and rate $\lambda > 0$, 
\begin{equation*}
 \pi\left(\alpha_{c}\right) = \lambda^{\gamma} \Gamma\left(\gamma\right)^{-1}\alpha_{c}^{\gamma-1}e^{-\lambda\alpha_{c}}, ~c=1,2,\dots, C,
\end{equation*}
where $\Gamma(\gamma) = \int_{0}^{\infty} t^{\gamma - 1} e^{-t} \text{d}t$. Similar to \citet{vitelliProbabilisticPreferenceLearning2017}, for the modal ranking we use a uniform prior on $\mathcal{P}_{m}$,
\begin{equation*}
\pi\left(\bm{\rho}_{c}\right) = \left(m!\right)^{-1} 1\left\{\bm{\rho}_{c} \in \mathcal{P}_{m}\right\} , ~ c= 1, 2, \dots, C.
\end{equation*}
For the cluster probabilities we use a symmetric Dirichlet prior,
\begin{equation*}
    \pi\left(\tau_{1}, \tau_{2}, \dots, \tau_{C}\right) = \Gamma\left(\psi C\right) \Gamma\left(\psi\right)^{-C} \prod_{c=1}^{C} \tau_{c}^{\psi-1}.
\end{equation*}

With non-transitive pairwise preferences, we use the Bernoulli model of \citet{crispinoBayesianMallowsApproach2019} with a truncated Beta prior on $[0, 0.5)$, 
\begin{equation*}
    \pi\left(\epsilon\right) \propto \epsilon^{\kappa_{1} - 1} \left(1 - \epsilon\right)^{\kappa_{2} - 1} 1\left\{\epsilon \in [0, 0.5)\right\}.
\end{equation*}
A model without non-transitive pairwise preferences corresponds to the limit $\kappa_{1} \to \infty$ and $\kappa_{2} \to 0$, which means that $\epsilon$ is fixed to $0$ and can be ignored in the analyses. \citet{crispinoBayesianMallowsApproach2019} also considered a logistic model, in which the logit of the error probability depends on the distance between the items in the latent ranking. While sequential inference with this logistic model is in principle straightforward, we do not consider it further in this paper for ease of presentation.

\subsection{Distance Functions and Normalizing Constants}
\label{sec:DistanceFunctions}

Consider two rankings $\bm{a}, \bm{b} \in \mathcal{P}_{m}$. Cayley distance measures the minimum number of pairwise swaps needed for converting $\bm{a}$ into $\bm{b}$ \citep{cayleyLXXVIINoteTheory1849}, two algorithms for which are given in \citet[pp. 25-26]{mardenAnalyzingModelingRank1995}. Ulam distance can be defined as $m$ minus the length of the longest common subsequence of the item orderings corresponding to $\bm{a}$ and $\bm{b}$ \citep{gordonMeasureAgreementRankings1979}. We further have Spearman distance $d(\bm{a}, \bm{b}) = \|\bm{a} - \bm{b}\|_{2}^{2}$ \citep{spearmanProofMeasurementAssociation1904}, Kendall distance measuring the number of discordant pairs $d(\bm{a}, \bm{b}) = \sum_{i=1}^{m} \sum_{j=i+1}^{m} 1{\{(a_{i} - a_{j}) (b_{i} - b_{j}) < 0\}}$ \citep{kendallNewMeasureRank1938}, the footrule $d(\bm{a}, \bm{b}) = \sum_{i=1}^{m} |a_{i} - b_{i}|$ \citep{spearmanFootruleMeasuringCorrelation1906}, and Hamming distance $d(\bm{a}, \bm{b}) = \sum_{i=1}^{m} 1\{a_{i} \neq b_{i}\}$ \citep{hammingErrorDetectingError1950}. 

The choice of distance in the Mallows model (\ref{eq:MallowsMixture}) is ultimately linked to the application at hand. For example, Hamming distance is not likely to work well under the ranking and preference applications considered in this paper but arises naturally when the Mallows model is used for matchings, e.g., when tracking a number of known objects using noisy sensors \citep{irurozkiMallowsGeneralizedMallows2019}. Similarly, Cayley distance is suitable when total disorder is of interest \citep{crispinoBayesianMallowsApproach2019}, whereas footrule, Kendall, and Spearman distance are most appropriate for preference data. \citet{crispinoBayesianMallowsApproach2019} gives an example of two rankings $\bm{a}=(1,2,3,4,5)$ and $\bm{b}=(5,2,3,4,1)$. Their Cayley distance (normalized to be in $[0,1]$) is 0.25, whereas the normalized footrule, Kendall, and Spearman distances are 2/3, 0.7, and 0.8, respectively. If $\bm{a}$ and $\bm{b}$ represent positions on a genome they can be seen as close, and the Cayley distance may be most appropriate. On the other hand, if $\bm{a}$ and $\bm{b}$ represent preferences for five items, they are far apart and one of the latter three distances are better. An in-depth comparison of Cayley, Kendall and Ulam distances can be found in \citet{ceberioReviewDistancesMallows2015} and further discussion in \citet[Ch. 6]{diaconisGroupRepresentationsProbability1988}.

Computing the normalizing constant $Z(\alpha)$ as in \eqref{eq:MallowsNormalizingConstant} requires summing over $\vert \mathcal{P}_{m} \vert = m!$ terms, but tractable exact expressions exist for Cayley, Kendall, and Hamming distances \citep{flignerDistanceBasedRanking1986,irurozkiSamplingLearningMallows2018}. Furthermore, since $d(\bm{r}, \bm{e})$ takes on a set of $l<m!$ values $\mathcal{D} = \{d_{1}, d_{2}, \dots, d_{l}\}$, we can define $L_{i} = \{\bm{r} \in \mathcal{P}_{m} : d(\bm{r}, \bm{e}) = d_{i}\}$ and write
$Z(\alpha) = \sum_{i=1}^{l} | L_{i}| e^{-\alpha d_{i}}$. For the footrule, $l = \mathcal{O}(m^{2})$, for Spearman distance $l = \mathcal{O}(m^{3})$, and for Ulam distance $l = \mathcal{O}(m)$  \citep{crispinoBayesianLearningMallows2018,crispinoEfficientAccurateInference2023,irurozkiSamplingLearningDistancebased2014,irurozkiPerMallowsPackageMallows2016}. Unfortunately, while the set of distances $\mathcal{D}$ is well known, finding the cardinalities $|L_{i}|$ is hard. The Online Encyclopedia of Integer Sequences \citep{sloaneEncyclopediaIntegerSequences2023} contains $|L_{i}|$ up to $m=50$ for the footrule, up to $m=20$ for Spearman distance, and $m=60$ for Ulam distance. Beyond these upper limits, asymptotic approximations exist for the footrule and Spearman distances \citep{crispinoEfficientAccurateInference2023,mukherjeeEstimationExponentialFamilies2016}, and an importance sampling scheme has been developed by \citet{vitelliProbabilisticPreferenceLearning2017}. The latter can in principle be run to arbitrary precision, and importantly, estimates of the partition function can be precomputed for a given number of items $m$ over a grid of $\alpha$ values \citep{sorensenBayesMallowsPackageBayesian2020}. Thus, in the rest of this paper we assume that $Z(\alpha)$ is available and that its Monte Carlo error (if any) is negligible compared to the Monte Carlo error of the proposed SMC$^{2}$ algorithm. In the simulations and application examples, the number of items $m$ is always such that $Z(\alpha)$ is known exactly.

\section{Sequential Inference in the Bayesian Mallows Model}
\label{sec:SequentialInference}

Now assume data become available sequentially at timepoints $t=1, 2, \dots, T$, and let $\bm{y}_{\mathcal{I}_{t}}$ contain partial rankings or pairwise preferences for new users entering the pool at time $t$, where $\mathcal{I}_{t} \subset \mathbb{N}$ is the set of user indices. Let $\mathcal{I}_{1:t} = \cup_{t=1}^{T} \mathcal{I}_{t}$ contain the indices of all users in the pool at time $t$. We assume throughout that a given user enters the pool only once, i.e., that 
$\mathcal{I}_s\cap\mathcal{I}_t = \emptyset$ if $s\neq t$. The target distribution is $\pi(\theta, \bm{x}_{\mathcal{I}_{1:t}} | \bm{y}_{\mathcal{I}_{1:t}})$, with static parameters $\theta = \{[\alpha_{c}, \bm{\rho}_{c}, \tau_{c}]_{c=1}^{C}, \epsilon\}$ and latent variables $\bm{x}_{\mathcal{I}_{1:t}} = \{\bm{r}_{\mathcal{I}_{1:t}}, z_{\mathcal{I}_{1:t}}\}$. The goal is to estimate the target distribution at all timepoints, in order to continuously perform inference based on the currently available evidence.

Sequential Monte Carlo (SMC) \citep{daiInvitationSequentialMonte2022,delmoralSequentialMonteCarlo2006,fearnheadParticleFiltersData2018,naessethElementsSequentialMonte2019} typically scales better than MCMC for these types of problems, as the latter needs to be completely rerun at each new timepoint. \citet{steinSequentialInferenceMallows2023} considered SMC for sequential inference in Bayesian Mallows models using a resample-move framework \citep{berzuiniRESAMPLEMOVEFilteringCrossModel2001,chopinSequentialParticleFilter2002,gilksFollowingMovingTarget2001}. Unfortunately, such methods were designed for cases either with only static parameters or latent variables which can be easily integrated out. Integrating over the latent variables in a Bayesian Mallows model, in particular the latent rankings, is computationally demanding and we thus instead base our methodology on SMC$^{2}$ \citep{chopinSMC2EfficientAlgorithm2013,fulopEfficientLearningSimulation2013} which uses particle marginal Metropolis-Hastings \citep{andrieuParticleMarkovChain2010} in the rejuvenation step and was developed specifically for settings with challenging latent variable distributions. We extend the SMC$^{2}$ framework by incorporating the hybrid particle MCMC sampler proposed by \citet{mendesFlexibleParticleMarkov2020} to allow a combination of Gibbs sampling and Metropolis-Hastings steps.

To set the notation, assume we have $R$ particles each containing static parameters $\theta^{r} = \{\alpha_{c}^{r}, \bm{\rho}_{c}^{r}, \tau_{c}^{r}, \epsilon^{r}\}$ and to each of these we attach $S$ additional particles containing the latent variables for the users entered up to timepoint $t$, $\bm{x}_{\mathcal{I}_{1:t}}^{s,r} = \{\bm{r}_{\mathcal{I}_{1:t}}^{s,r}, z_{\mathcal{I}_{1:t}}^{s,r}\}$. Here and in what follows, superscripts $r$ and $s$ are assumed repeated for all $r=1,2,\dots,R$ and $s=1,2,\dots,S$ and subscripts $c$ are assumed repeated for $c=1,2,\dots,C$. If the data consist of either partial rankings or consistent pairwise preferences we have $\epsilon^{r}=0$ and this parameter can be ignored. Similarly, in the absence of mixtures  $\tau_{c}^{r}=1$ and $z_{i}^{s,r}=1$ are fixed and can be ignored.

The building blocks of the algorithm are particle filters for latent rankings and cluster labels (Section \ref{sec:ParticleFilter}), iterated batch importance sampling for the static parameters (Section \ref{sec:SMCSquared}), and a rejuvenation algorithm (Section \ref{sec:RejuvenationAlgorithm}). Proposal distributions for latent rankings are discussed in Section \ref{sec:LatentRankingProposal}.

\subsection{Particle Filters for Latent Rankings}
\label{sec:ParticleFilter}

Define ancestor indices $a_{t-1}^{s}$ indicating which particle at time $t-1$ is the ancestor of particle $s$ at time $t$, and set the initial value $a_{0}^{s}=s$. At timepoint $t$, $\bm{x}_{n,t-1}^{a_{t-1}^{s}}$ denotes the latent variables of user $n \in \mathcal{I}_{1:t-1}$ in particle $s$. The $r$ superscripts linking latent variables to static parameter particles are omitted for ease of notation. Also let $\mathcal{M}_{C}(\bm{p})$ denote a multinomial distribution over $C \in \mathbb{N}$ categories with probabilities $\bm{p}$. Algorithm \ref{alg:BasicParticleFilter} approximates $\pi_{\theta}(\bm{x}_{I_{1:T}}) = p(\bm{x}_{I_{1:T}} | \bm{y}_{I_{1:T}}, \theta)$ for fixed $\theta$.

\begin{algorithm}[htb]
\caption{Particle Filter}
\label{alg:BasicParticleFilter}
\begin{algorithmic}[1]
\For{$t = 1$ to $T$}
    \If{$t > 1$}
        \State Sample $a_{t-1}^{s} \in [S]$ with probabilities $W_{t-1,\theta}^{1:S}$.
        \For{$n \in \mathcal{I}_{1:t-1}$}
            \State $\mathbf{x}_{n,t}^{s} \gets \mathbf{x}_{n,t-1}^{a_{t-1}^{s}}$.
        \EndFor
    \EndIf
    \For{$n \in \mathcal{I}_{t}$}
        \State Sample $\bm{r}_{n,t}^{s} \sim q_{\theta}(\cdot | \mathcal{S}_{n})$.
        \State Sample $z_{n,t}^{s} \sim \mathcal{M}_{C}(\bm{p}_{n})$ with probabilities
        \begin{equation}
        \label{eq:BayesMallowsClusterProbs}
        p_{n,c} = \frac{\tau_{c} Z(\alpha_{c})^{-1} \exp\{-\alpha_{c} d(\bm{r}_{n,t}^{s}, \bm{\rho}_{c})\}}{\sum_{c=1}^{C}\tau_{c} Z(\alpha_{c})^{-1} \exp\{-\alpha_{c} d(\bm{r}_{n,t}^{s}, \bm{\rho}_{c})\}}.
        \end{equation}
    \EndFor
    \State Compute weights
    \begin{align}
    \label{eq:ParticleFilterWeights}
    &w_{t,\theta}^{s} = \\
    \nonumber
    & \prod_{n \in \mathcal{I}_{t}} \frac{\sum_{c=1}^{C}\tau_{c} Z(\alpha_{c})^{-1} \exp\{-\alpha_{c} d(\bm{r}_{n,t}^{s}, \bm{\rho}_{c})\}}{q_{\theta}(\bm{r}_{n,t}^{s} | \mathcal{S}_{n})} \left(\frac{\epsilon}{1 - \epsilon}\right)^{\sum_{i=1}^{p_{n,t}} g\left(y_{ni}, \bm{r}_{n,t}^{s}\right)} \left(1-\epsilon\right)^{p_{n,t}}.
    \end{align}
    \State Normalize weights 
    \begin{align}
    \label{eq:NormalizeWeights}
        W_{t,\theta}^{s} = \frac{w_{t,\theta}^{s}}{\sum_{s=1}^{S} w_{t,\theta}^{s}}.
    \end{align}
\EndFor
\end{algorithmic}
\end{algorithm}

In Algorithm \ref{alg:BasicParticleFilter} the loop on lines 4-6 ensures that estimated latent rankings for all users are available at each timepoint, but can be omitted to reduce the memory cost. On line 9, $\mathcal{S}_{n}$ is given by \eqref{eq:ConstraintSetPartial} in the case of partial rankings, \eqref{eq:ConstraintSetPairwise} in the case of consistent pairwise preferences, and $\mathcal{P}_{m}$ in the case of non-transitive pairwise preferences. We postpone the details of these proposal distributions to Section \ref{sec:LatentRankingProposal}.

The expression for the weights \eqref{eq:ParticleFilterWeights} is based on Step 2(c) of \citet[Sec. 2.1]{chopinSMC2EfficientAlgorithm2013} which in our notation becomes
\begin{equation*}
    w_{t,\theta}^{s} = \frac{f_{\theta}\left(\bm{x}_{\mathcal{I}_{t}}^{s} | \bm{x}_{\mathcal{I}_{1:t-1}}^{a_{t-1}^{s}}\right) p_{\epsilon}\left(\bm{y}_{\mathcal{I}_{t}} | \bm{x}_{\mathcal{I}_{t}}^{s}\right)}{q_{t,\theta}\left(\bm{x}_{\mathcal{I}_{t}}^{s} | \bm{x}_{\mathcal{I}_{t-1}}^{a_{t-1}^{s}}\right)}.
\end{equation*}
Because independent users arrive at each timepoint, $\bm{x}_{\mathcal{I}_{t}}^{s}$ and $\bm{x}_{\mathcal{I}_{1:t-1}}^{a_{t-1}^{s}}$ are independent given $\theta$ and we get
\begin{align*}
    f_{\theta}\left(\bm{x}_{\mathcal{I}_{t}}^{s} | \bm{x}_{\mathcal{I}_{1:t-1}}^{a_{t-1}^{s}} \right) &= f_{\theta}\left(\bm{x}_{\mathcal{I}_{t}}^{s} \right) 
    = f_{\theta}\left(z_{\mathcal{I}_{t}}^{s}\right) f_{\theta}\left(\bm{r}_{\mathcal{I}_{t}}^{s} | z_{\mathcal{I}_{t}}^{s} \right) \\
    &= \prod_{n \in \mathcal{I}_{t}}\frac{\tau_{z_{n,t}^{s}} }{Z\left(\alpha_{z_{n,t}^{s}}\right)} \exp\left\{-\alpha_{z_{n,t}^{s}} d\left(\bm{r}_{n,t}^{s}, \bm{\rho}_{z_{n,t}^{s}}\right) \right\}.
\end{align*}
Next, it follows from Section \ref{sec:MallowsBackground} that 
\begin{equation*}
    p_{\epsilon}\left(\bm{y}_{\mathcal{I}_{t}} | \bm{x}_{\mathcal{I}_{t}}^{s}\right) =  \prod_{n \in \mathcal{I}_{t}} \left(\frac{\epsilon}{1 - \epsilon}\right)^{\sum_{i=1}^{p_{n,t}} g\left(y_{ni}, \bm{r}_{n,t}^{s}\right)} \left(1-\epsilon\right)^{p_{n,t}}
\end{equation*}
which simplifies to $p_{\epsilon}(\bm{y}_{\mathcal{I}_{t}} | \bm{x}_{\mathcal{I}_{t}}^{s}) = 1$ when $\epsilon = 0$ and where $p_{n,t}$ denotes the number of pairwise preferences in $\bm{y}_{n}$ for some $n \in \mathcal{I}_{t}$. The proposal distribution is
\begin{align*}
    q_{t,\theta}\left(\bm{x}_{\mathcal{I}_{t}}^{s} | \bm{x}_{\mathcal{I}_{t-1}}^{a_{t-1}^{s}}\right) &= \prod_{n \in \mathcal{I}_{t}}q_{\theta}\left(\bm{r}_{n,t}^{s} | \mathcal{S}_{n}\right) q_{\theta}\left(z_{n,t}^{s} | \bm{r}_{n,t}^{s}\right) \\
    &= \prod_{n \in \mathcal{I}_{t}} q_{\theta}\left(\bm{r}_{n,t}^{s} | \mathcal{S}_{n}\right) \frac{\tau_{z_{n,t}^{s}} Z\left(\alpha_{z_{n,t}^{s}}\right)^{-1} \exp\left\{-\alpha_{z_{n,t}^{s}}d\left(\bm{r}_{n,t}^{s}, \bm{\rho}_{z_{n,t}^{s}}\right) \right\}}{\sum_{c=1}^{C}\tau_{c} Z\left(\alpha_{c}\right)^{-1} \exp\left\{-\alpha_{c}d\left(\bm{r}_{n,t}^{s}, \bm{\rho}_{c}\right) \right\}} .
\end{align*}
These three expression combine to yield the fraction in \eqref{eq:ParticleFilterWeights}. For the special case $C=1$ and $\epsilon=0$ we recover the weight update formula in Algorithm 12 of \citet{steinSequentialInferenceMallows2023} and with complete rankings we recover Algorithm 14 of \citet{steinSequentialInferenceMallows2023}.

In the resampling step on line 3 in Algorithm \ref{alg:BasicParticleFilter}, as well as in the resampling steps of all subsequent algorithms, both multinomial resampling \citep{gordonNovelApproachNonlinear1993} and the lower variance alternatives residual resampling \citep{liuSequentialMonteCarlo1998}, stratified resampling \citep{kitagawaMonteCarloFilter1996}, and systematic resampling \citep{kitagawaMonteCarloFilter1996} can be used and are part of our implementation. We refer to \citet{doucComparisonResamplingSchemes2005} and \citet{holResamplingAlgorithmsParticle2006} for details. 

We also note the important fact that the quantity
\begin{equation}
\label{eq:Zhat}
    \hat{Z}_{t}\left(\theta, \bm{x}_{\mathcal{I}_{1:t}}^{1:S}, a_{1:t-1}^{1:S}\right) = \frac{1}{S^{t}}  \prod_{t'=1}^{t} \left\{ \sum_{s=1}^{S} w_{t',\theta}^{s}\right\}
\end{equation}
is an unbiased estimator of the marginal likelihood $p(\bm{y}_{\mathcal{I}_{1:t}}|\theta)$ \citep[Sec. 7.4.1]{delmoralFeynmanKacFormulae2004}.

\subsubsection{Conditional Particle Filter}

To allow particle Gibbs sampling in the rejuvenation step, we need a conditional particle filter \citep{andrieuParticleMarkovChain2010} for which the full ancestral history of a given particle $\bm{x}_{\mathcal{I}_{1:T}}^{k}$ is fixed. This particle filter is shown in Algorithm \ref{alg:ConditionalPF} and yields samples approximately distributed according to $p(\bm{x}_{\mathcal{I}_{1:T}}^{-k} | \bm{y}_{\mathcal{I}_{1:T}}, \theta, \bm{x}_{\mathcal{I}_{1:T}}^{k})$, where $k \in \{1,2,\dots,S\}$ and $\bm{x}_{\mathcal{I}_{1:T}}^{-k}$ denotes the set of all particles except particle $k$. 

\begin{algorithm}[htb]
\caption{Conditional Particle Filter}
\label{alg:ConditionalPF}
\begin{algorithmic}[1]
\State Condition on a trajectory $\bm{x}_{\mathcal{I}_{1:T}}^{k}$ with ancestral lineage $b_{T}^{k}=k$ and $b_{t}^{k} = a_{t}^{b_{t+1}^{k}}$, $t=T-1,\dots,1$.
\For{$t = 1$ to $T$}
    \If{$t > 1$}
        \State For $s \neq b_{t}^{k}$ sample $a_{t-1}^{s} \in [S]$ with probabilities $W_{t-1,\theta}^{1:S}$.
        \For{$n \in \mathcal{I}_{1:t-1}$}
            \State Set $\bm{x}_{n,t}^{s} \gets \bm{x}_{n,t-1}^{a_{t-1}^{s}}$.
        \EndFor
    \EndIf    
    \For{$n \in \mathcal{I}_{t}$}        
        \State For $s \neq b_{t}^{k}$ sample $\bm{r}_{n,t}^{s} \sim q_{\theta}(\cdot | \mathcal{S}_{n,t})$.
        \State For $s \neq b_{t}^{k}$ sample $z_{n,t}^{s} \sim \mathcal{M}(C)$ with probabilities \eqref{eq:BayesMallowsClusterProbs}.        
    \EndFor    
    \State Compute weights using \eqref{eq:ParticleFilterWeights} and normalize them using \eqref{eq:NormalizeWeights}.
\EndFor
\end{algorithmic}
\end{algorithm}

\subsection[placeholder]{SMC$^{2}$ Algorithm}
\label{sec:SMCSquared}

The top-level algorithm for sampling the static parameters is an extension of iterated batch importance sampling \citep{chopinSequentialParticleFilter2002} which uses the particle filters of the previous section to integrate out the latent variables, and is stated in Algorithm \ref{alg:SMC2}. Since the particle filters yield unbiased estimates of the marginal likelihood, Algorithm \ref{alg:SMC2} targets the correct posterior distribution $\pi(\theta | \bm{y}_{\mathcal{I}_{1:t}})$ at each $t=1,2,\dots,T$ \citep{chopinSMC2EfficientAlgorithm2013}.

\begin{algorithm}[htb]
\caption{SMC$^{2}$ Algorithm}
\label{alg:SMC2}
\begin{algorithmic}[1]
\State Sample $\theta^{r} = \{\alpha_{c}^{r},\bm{\rho}_{c}^{r},\tau_{c}^{r},\epsilon^{r}\}$ from their priors and set $\omega^{r} \gets 1$.
\For{$t = 1$ to $T$}
    \State Perform iteration $t$ of the particle filter in Algorithm \ref{alg:BasicParticleFilter} with $\theta=\theta^{r}$ and compute
    \begin{equation}
    \label{eq:SMC2incremental}
    \hat{p}\left(\bm{y}_{\mathcal{I}_{t}} | \bm{y}_{\mathcal{I}_{1:t-1}}, \theta^{r}\right) = \frac{1}{S}\sum_{s=1}^{S} w_{t,\theta}^{s}.
    \end{equation}    
    \State Update and normalize importance weights
    \begin{equation}
    \label{eq:SMC2WeightUpdate}
    \omega^{r} \gets \omega^{r} \times \hat{p}\left(\bm{y}_{\mathcal{I}_{t}} | \bm{y}_{\mathcal{I}_{1:t-1}}, \theta^{r}\right), \quad \Omega^{r} = \frac{\omega^{r}}{\sum_{r=1}^{R} \omega^{r}}.
    \end{equation}    
    \State Compute the effective sample size $\text{ESS} = \{\sum_{r=1}^{R} (\Omega^{r})^{2}\}^{-1}$.    
    \If{$\text{ESS} < A$}
        \State Sample $a_{t}^{r} \in [R]$ with probabilities $\Omega^{r}$ and set $(\theta^{r}, \omega^{r}) \gets (\theta^{a_{t}^{r}}, 1)$.
        \State Rejuvenate with Algorithm \ref{alg:Rejuvenation}, letting $\zeta$ denote the acceptance rate of \eqref{eq:RejuvenationMHRatio}.
        \If{$\zeta < B$} 
            \State Set $\tilde{S}=2S$ and sample $i^{\tilde{s}} \in [S]$ for $\tilde{s}=1,\dots,\tilde{S}$ with probabilities $W_{t,\theta^{r}}^{s}$.            
            \State Set $\{\tilde{\bm{x}}_{\mathcal{I}_{1:t}}^{1:\tilde{S}}, \tilde{a}_{1:t-1}^{1:\tilde{S}}\} \gets \{\bm{x}_{\mathcal{I}_{1:t}}^{i^{1:\tilde{S}}}, a_{1:t-1}^{i^{1:\tilde{S}}}\}$ and $\tilde{w}_{1:t,\theta^{r}}^{1:\tilde{S}} \gets w_{1:t,\theta^{r}}^{i^{1:\tilde{S}}}$.            
            \State Update $S \gets \tilde{S}$ and the particle weight
            \begin{equation}
            \label{eq:FilterIncreaseUpdate}
            \omega^{r} \gets \omega^{r} \times \left(S / \tilde{S}\right)^{t} \frac{  \prod_{t'=1}^{t} \left\{ \sum_{s=1}^{\tilde{S}} \tilde{w}_{t,\theta^{r}}^{s}\right\}}{\prod_{t'=1}^{t} \left\{ \sum_{s=1}^{S} w_{t,\theta^{r}}^{s}\right\}}.
            \end{equation}            
        \EndIf
    \EndIf
\EndFor
\end{algorithmic}
\end{algorithm}

Considering Algorithm \ref{alg:SMC2}, first note that equation \eqref{eq:SMC2incremental} is an unbiased estimator of
\begin{equation}
\label{eq:CompleteRankingsIncrement}
    p\left(\bm{y}_{\mathcal{I}_{t}} | \bm{y}_{\mathcal{I}_{1:t-1}}, \theta^{r}\right) = \prod_{n \in \mathcal{I}_{t}}\sum_{c=1}^{C} \tau_{c}^{r} Z\left(\alpha_{c}^{r}\right)^{-1} \sum_{\bm{r}_{n} \in \mathcal{S}_{n}} \exp\left\{-\alpha_{c}^{r} d\left(\bm{r}_{n}, \bm{\rho}_{c}^{r}\right)\right\}.
\end{equation}
The marginal likelihood increments are given by 
\begin{equation}
\label{eq:MarginalLikelihoodNewObs}
    \hat{p}\left(\bm{y}_{\mathcal{I}_{t}} | \bm{y}_{\mathcal{I}_{1:t-1}}\right) = \sum_{r=1}^{R} \Omega^{r} \times \hat{p}\left(\bm{y}_{\mathcal{I}_{t}} | \bm{y}_{\mathcal{I}_{1:t-1}}, \theta^{r}\right)
\end{equation}
and can be used to estimate the unconditional marginal likelihood
\begin{equation}
\label{eq:SMC2MarginalLikelihood}
    \hat{p}\left(\bm{y}_{\mathcal{I}_{1:t}}\right) = \prod_{t'=1}^{t} \hat{p}\left(\bm{y}_{\mathcal{I}_{t'}} | \bm{y}_{\mathcal{I}_{1:t'-1}}\right).
\end{equation}

The rejuvenation threshold $A$ can be set to $R/2$. As in \citet{fulopEfficientLearningSimulation2013}, we iterate the rejuvenation algorithm at least once, and stop when the number of unique particles exceeds $A$ or when some upper limit on the number of iterations is reached. If the acceptance rate in the rejuvenation step is below some threshold $B$, which we set to $B=0.2$ here, the number of particle filters is doubled. The doubling on lines 10-12 implements the exchange importance sampling step of \citet[Sec. 3.6.1]{chopinSMC2EfficientAlgorithm2013}. The components in \eqref{eq:FilterIncreaseUpdate} are readily available from the call to Algorithm \ref{alg:Rejuvenation} in the rejuvenation step.

\subsubsection{Parallelization}

To reduce the amount of communication between nodes, it seems most sensible to parallelize the top-level Algorithm \ref{alg:SMC2} rather than the particle filters. There are two main approaches to this in the literature \citep{daiInvitationSequentialMonte2022,naessethElementsSequentialMonte2019}.  \citet{junEntangledMonteCarlo2012} and \citet{murrayParallelResamplingParticle2016} use the fact that the weight updates in equations \eqref{eq:SMC2incremental}-\eqref{eq:SMC2WeightUpdate} can be done independently for each of the $R$ particles. However, computing effective sample size and subsequently resampling requires communication between the nodes, and hence makes its implementation complicated. 

A more straightforward approach, which we use in this paper, is what \citet{naessethElementsSequentialMonte2019} call importance weighted SMC samplers. In this case the full algorithm with $R$ particles is run independently on $P$ different nodes. Let $\theta^{r,p}$ denote the $r$th particle of the SMC$^{2}$ algorithm run on the $p$th compute node and $\Omega^{r,p}$ its weight, for $r=1,\dots,R$ and $p=1,\dots,P$. Also let $\hat{p}(\bm{y}_{\mathcal{I}_{1:T}})^{p}$ denote the marginal likelihood estimate \eqref{eq:SMC2MarginalLikelihood} from the $p$'th node. The combined set of particles $\{\theta^{r,p}\}$ with weights
\begin{equation}
\label{eq:CombineParallel}
    \Omega^{r,p} \times \frac{\hat{p}(\bm{y}_{\mathcal{I}_{1:T}})^{p}}{\sum_{p'=1}^{P}\hat{p}(\bm{y}_{\mathcal{I}_{1:T}})^{p'}}
\end{equation}
now yield a consistent estimate of the target distribution as $P \to \infty$ for any $R$ \citep[Sec. 4.4.1]{naessethElementsSequentialMonte2019}. The combined estimate of the marginal likelihood itself can be obtained by direct averaging, $\hat{p}(\bm{y}_{\mathcal{I}_{1:T}}) = \sum_{p'=1}^{P} \hat{p}(\bm{y}_{\mathcal{I}_{1:T}})^{p'} /P$.

\subsubsection{Latent Variable Prediction}

Latent variable prediction corresponds to state inference in the SMC context. Predicting the latent variables $\bm{x}_{\mathcal{I}_{t}} = \{\bm{r}_{\mathcal{I}_{t}}, z_{\mathcal{I}_{t}}\}$ of the users entering at time $t$ is a filtering problem, and we can obtain $R$ samples, weighted by $\Omega^{r}$, from $p(\bm{x}_{\mathcal{I}_{t}} | \theta, \bm{y}_{\mathcal{I}_{1:t}})$ by drawing an index $s \sim \mathcal{M}(W_{t,\theta}^{r,s})$ for each particle $r$ \citep[Sec. 3.3]{chopinSMC2EfficientAlgorithm2013}. 

Sampling from the posterior $P(\bm{x}_{\mathcal{I}_{1:t}} | \theta, \bm{y}_{\mathcal{I}_{1:t}})$ of the latent variables of all users entered until time $t$ can be done identically, but requires storing the full path for each particle. That is, if we draw a particle with index $s$ we need to trace its latent variables according to its genealogy back until time 1. As noted by \citet{chopinSMC2EfficientAlgorithm2013}, this storage requirement can be avoided by triggering the particle doubling step in Algorithm \ref{alg:SMC2} whenever a complete trajectory of the latent variables are needed, and then sampling an index $s \sim \mathcal{M}(W_{t,\theta}^{r,s})$ for each $r \in [R]$.

\subsection{Rejuvenation}
\label{sec:RejuvenationAlgorithm}

The rejuvenation step prevents degeneracy by moving each particle independently with an MCMC kernel. The original SMC$^{2}$ rejuvenation algorithms of \citet{chopinSMC2EfficientAlgorithm2013} and \citet{fulopEfficientLearningSimulation2013} used particle marginal Metropolis-Hastings, but we instead use the algorithm proposed in \citet{mendesFlexibleParticleMarkov2020} which combines particle marginal Metropolis-Hastings with particle Gibbs. This is useful in the present case because cluster probabilities $\tau_{c}$ and the error probability $\epsilon$ can be sampled conditionally, whereas the dispersion parameters $\alpha_{c}$ and the modal rankings $\bm{\rho}_{c}$ require a Metropolis-Hastings algorithm. For ease of notation, now let $T$ denote the current value $t$ of SMC$^{2}$ at the moment the rejuvenation algorithm is called. Also define $a \land b = \text{min}\{a, b\}$. 

\begin{algorithm}[H]
\caption{Rejuvenation Algorithm}
\label{alg:Rejuvenation}
\begin{algorithmic}[1]
\State Compute $\hat{\sigma}_{\alpha,c}^{2} = \frac{1}{R} \sum_{r=1}^{R} (\alpha_{c}^{r} -  \hat{\alpha}_{c})^2$ where $\hat{\alpha}_{c} = \frac{1}{R}\sum_{r=1}^{R}\alpha_{c}^{r}$.
\State Sample $k \in [S]$ with probabilities $W_{T,\theta^{r}}^{1:S}$.
\While{stopping criterion not met}    
    \State Sample proposals $\alpha_{c}' \sim \log\mathcal{N}\left(\log \alpha_{c}^{r}, \hat{\sigma}_{\alpha,c}^{2}\right)$ and $\bm{\rho}_{c}' \sim LS\left(\bm{\rho}_{c}^{r}\right)$ and set $\theta' \gets \{\alpha_{c}', \bm{\rho}_{c}', \tau_{c}, \epsilon\}_{c=1}^{C}$.        
    \State Run a particle filter (Algorithm \ref{alg:BasicParticleFilter}) for $t=1,2,\dots,T$ and compute
    \begin{equation*}    
    \hat{Z}_{T}\left(\theta', \bm{x}_{\mathcal{I}_{1:T}}^{1:S}, a_{1:T-1}^{1:S}\right) = \prod_{t=1}^{T} \left\{\frac{1}{S} \sum_{s=1}^{S} w_{t,\theta'}^{s}\right\}.
    \end{equation*}
    \State Sample $k' \in [S]$ with probabilities $W_{T,\theta'}^{1:S}$ (from particle filter).
    \State Set $(\theta^{r}, k) \gets (\theta', k')$ with probability
    \begin{equation}
    \label{eq:RejuvenationMHRatio}
    1 \land \frac{\hat{Z}_{T}\left(\theta', \bm{x}_{\mathcal{I}_{1:T}}^{1:S}, a_{1:T-1}^{1:S}\right)}{\hat{Z}_{T}\left(\theta^{r}, \bm{x}_{\mathcal{I}_{1:T}}^{1:S}, a_{1:T-1}^{1:S}\right)} \prod_{c=1}^{C} \left(\frac{\alpha_{c}'}{\alpha_{c}^{r}}\right)^{\gamma} \exp\left\{-\lambda \left(\alpha_{c}' - \alpha_{c}^{r}\right)\right\} .
    \end{equation}
    \State Define $\bm{x}_{\mathcal{I}_{1:t}}^{k}=\{\bm{r}_{n}^{k}, z_{n}^{k}\}_{n \in \mathcal{I}_{1:t}}$ as the latent variables in particle filter $k$.
    \State Compute $\hat{N}_{c}=\sum_{n \in \mathcal{I}_{1:T}} 1\{z_{n}^{k} = c\}$ and $\hat{\psi}_{c} = \psi + \hat{N}_{c}$, and sample
    \begin{equation*}
        \bm{\tau}' \sim \text{Dirichlet}\left(\hat{\psi}_{1}, \hat{\psi}_{2}, \dots, \hat{\psi}_{C}\right) = \Gamma\left(\sum_{c=1}^{C}\hat{\psi}_{c}\right)\left\{\prod_{c=1}^{C} \Gamma\left(\hat{\psi}_{c}\right)\right\}^{-1} \prod_{c=1}^{C} \tau_{c}^{\hat{\psi}_{c}-1}.
    \end{equation*}
    \State Compute $a=\sum\limits_{n \in \mathcal{I}_{1:t}} \sum_{i=1}^{p_{n}} g(y_{ni}, \bm{r}_{n}^{k})$, $b=\sum\limits_{n \in \mathcal{I}_{1:t}} \sum_{i=1}^{p_{n}}[1- g(y_{ni}, \bm{r}_{n}^{k})]$, and sample
    \begin{equation*}
        \epsilon' \sim f(\epsilon)  \propto \epsilon^{\kappa_{1} -1+a } (1 - \epsilon)^{\kappa_{2} -1+ b } 1\{ \epsilon \in [0, 0.5)\}.
    \end{equation*}    
    \State Set $\theta^{r} \gets \{\alpha_{c}^{r}, \bm{\rho}_{c}^{r}, \tau_{c}', \epsilon'\}_{c=1}^{C}$.        
    \State Run Algorithm \ref{alg:ConditionalPF}, conditional on $\bm{x}_{\mathcal{I}_{1:T}}^{k}$, for $t=1,2,\dots,T$.     
    \State Sample $k \in [S]$ with probabilities $W_{T,\theta^{r}}^{1:S}$ (from conditional particle filter).    
\EndWhile
\end{algorithmic}
\end{algorithm}

Algorithm \ref{alg:Rejuvenation} starts by computing the variance of each $\alpha_{c}$, using unweighted formulas because we always resample before rejuvenating. These variances are used for tuning the random walk proposal on line 4. Next, on line 2, we sample a complete particle history $\bm{x}_{\mathcal{I}_{1:T}}^{k}$ from the particle filters previously run with the parameter value $\theta$. On line 4, $LS(\cdot)$ denotes the leap-and-shift proposal defined in Algorithm \ref{alg:LeapAndShift}. To avoid introducing another tuning parameter, and to keep the proposal symmetric, we set leap size to 1. The extension to larger leap sizes is straightforward, and we refer to \citet[Sec. 2.4]{vitelliProbabilisticPreferenceLearning2017} for details. 

\begin{algorithm}[htb]
\caption{Leap-and-Shift Proposal for Modal Ranking \citep{vitelliProbabilisticPreferenceLearning2017}}
\label{alg:LeapAndShift}
\begin{algorithmic}[1]
\Input{The current value $\bm{\rho}$.}
\Output{A proposal $\bm{\rho}'$ separated from $\bm{\rho}$ by an Ulam distance of 1.}
\State Sample uniformly $u \sim \mathcal{U}\{1,\dots,m\}$.
\State Define $\mathcal{S} = \{\text{max}(1, \rho_{u} - 1), \text{min}(m, \rho_{u}+1)\} \setminus \{\rho_{u}\}$.
\State Sample uniformly $r \sim \mathcal{U}\{\mathcal{S}\}$.
\State Define $\bm{\rho}^{*} \in \{1,\dots,m\}^{m}$ with elements $\rho_{u}^{*}=r$ and $\rho_{i}^{*}=\rho_{i}$ for $i \in \{1, \dots, m\} \setminus \{u\}$.
\State Define $\Delta = \rho_{u}^{*} - \rho_{u}$ and the proposal $\bm{\rho}' \in \mathcal{P}_{m}$ with elements
\begin{equation*}
    \rho_{i}' = 
    \begin{cases}
        \rho_{u}^{*} & \text{if } \rho_{i} = \rho_{u} \\
        \rho_{i} - 1 & \text{if } \rho_{u} < \rho_{i} \leq \rho_{u}^{*} \text{ and } \Delta > 0 \\ 
        \rho_{i} + 1 & \text{if } \rho_{u} > \rho_{i} \geq \rho_{u}^{*} \text{ and } \Delta < 0 \\
        \rho_{i} & \text{otherwise,}
    \end{cases}
\end{equation*}
for $i=1,\dots,m$.
\end{algorithmic}
\end{algorithm}

On line 5 a new particle filter is run in order to compute the marginal likelihood of the proposed parameters. On line 6 we sample a proposal $k'$ for a new particle history to condition on, using the weights from the particle filter run on line 5. The proposals $\theta'$ and $k'$ are accepted with probability given by the Metropolis-Hastings ratio \eqref{eq:RejuvenationMHRatio} in which the product term follows directly from the priors.

On line 9 we first compute the cluster frequencies $\hat{N}_{c}$ in the particle filter $k$ that we condition on, after which we sample the cluster probabilities from their conditional posterior \citep[Sec. 4.3]{vitelliProbabilisticPreferenceLearning2017}. On line 10 we sample the error probability from its conditional posterior \citep[p. 504]{crispinoBayesianMallowsApproach2019}. Lines 12 and 13, which consist of running a conditional particle filter and sampling a new $k \in [S]$ are necessary for computing the denominator in \eqref{eq:RejuvenationMHRatio} in the next iteration.

A reasonable stopping criterion which is computationally easy to check is that the number of unique parameters exceeds some threshold, say $R/2$. However, the Gibbs sampler is guaranteed to produce new values of all $\tau_{c}^{r}$ and hence when $C>1$ we are guaranteed to have $R$ unique particles after a single iteration of the algorithm. The same applies to $\epsilon^{r}$ when we have non-transitive pairwise preferences. If this is sufficient, lines 12-13 can be skipped and no conditional particle filter needs to be run. On the other hand, this may lead to degeneracy in $\alpha_{c}$ and $\bm{\rho}_{c}$, so in this case it might be more useful to monitor to the number of unique values of $\alpha_{c}^{r}$, and stop the rejuvenation when this number exceeds $R/2$.

\subsection{Proposals for Latent Rankings}
\label{sec:LatentRankingProposal}

In the particle filters of Section \ref{sec:ParticleFilter} we use a proposal for the latent rankings on the form $q_{\theta}(\cdot | \mathcal{S}_{n})$, where $\mathcal{S}_{n}$ is the set of rankings $\bm{r} \in \mathcal{P}_{m}$ compatible with the preferences given by user $n$. With partial rankings, $\mathcal{S}_{n}$ is given by \eqref{eq:ConstraintSetPartial} and with consistent pairwise preferences it is given by \eqref{eq:ConstraintSetPairwise}. We now consider these cases in turn. With non-transitive pairwise preferences we have $\mathcal{S}_{n} = \mathcal{P}_{m}$ and sampling proposals amounts to simply permuting the integers $[m]$, all of which have probability $1/m!$, and hence no further consideration needs to be given to this case.

\subsubsection{Partial Rankings}

The simplest approach, used by \citet{vitelliProbabilisticPreferenceLearning2017}, is to randomly permute the elements of $\mathcal{S}_{i}$ which are not fixed to a given rank. In this case the proposal distribution takes the form $q_{\theta}\left(\bm{r}_{n} | \mathcal{S}_{n} \right) = | \mathcal{S}_{n}|^{-1} 1\{ \bm{r}_{n} \in \mathcal{S}_{i} \}$
and is independent of $\theta$. Note that while this uniform distribution cancels out from the normalized weight formula \eqref{eq:NormalizeWeights}, the probability $1/|\mathcal{S}_{n}|$ needs to be explicitly added to the unnormalized weights in \eqref{eq:ParticleFilterWeights} for the marginal likelihood computation in \eqref{eq:MarginalLikelihoodNewObs} and subsequently \eqref{eq:SMC2MarginalLikelihood} to be correct.

\citet{steinSequentialInferenceMallows2023} developed an alternative pseudolikelihood proposal which uses information in $\theta$ when proposing a new partial ranking for the user. We present it in Algorithm \ref{alg:Pseudolikelihood}. For a given user $n$, it first fixes the observed items $\mathcal{A}_{n}$ to their given value and then iterates through the unranked elements $\mathcal{A} \setminus \mathcal{A}_{n}$ in random order, sampling conditionally on the hitherto realized ranks. The distance $d(\cdot, \cdot)$ used in \eqref{eq:PseudolikelihoodSampling} needs to be either footrule or Spearman, since only these have a natural definition between single elements of ranking vectors, but note that the Mallows model can use any of the distance functions discussed in Section \ref{sec:DistanceFunctions}. The key difference from a uniform proposal is that the distribution for a latent rank $r_{ni}$ in \eqref{eq:PseudolikelihoodSampling} is designed such that values close to the current estimate of the modal ranking $\rho_{i}$ for item $A_{i}$ are more likely to be obtained than values far from the modal ranking.

Pseudolikelihood proposal does not work for mixture models, as this would require knowledge of the cluster label $z_{n,t}^{s}$ for the given user in order to pick the right parameters $\alpha_{z_{n,t}^{s}}$ and $\bm{\rho}_{z_{n,t}^{s}}$. Since $z_{n,t}^{s}$ needs to be sampled after $\bm{r}_{n,t}^{s}$ in the particle filters, it is not directly clear how to achieve this.

\begin{algorithm}[htb]
\caption{Pseudolikelihood Proposal for Latent Rankings \citep{steinSequentialInferenceMallows2023}}
\label{alg:Pseudolikelihood}
\begin{algorithmic}[1]
\Input{Parameters $\theta = \{\bm{\alpha},\bm{\rho}\}$ and data $\bm{y}_{n}$.}
\Output{A proposal $\bm{r}_{n}$ and its probability $q_{\theta}(\bm{r}_{n} | \mathcal{S}_{n})$.}
\State Define $\mathcal{B}_{n} = \emptyset$, and $q_{\theta}(\bm{r}_{n} | \mathcal{S}_{n}) = 1$.
\For{$i : A_{i} \in \mathcal{A}_{n}$}
\State Set $r_{ni} = y_{ni}$ and $\mathcal{B}_{n} \gets \mathcal{B}_{n} \cup r_{ni}$.
\EndFor
\State Randomize the order of unranked items, $\bm{o}_{n} = \text{Permutation}(\mathcal{A} \setminus \mathcal{A}_{n})$.
\For{$i : A_{i} \in \bm{o}_{n}$}
\State Sample $r_{ni} \in [m] \setminus \mathcal{B}_{n}$ with probability 
\begin{equation}
\label{eq:PseudolikelihoodSampling}
 p\left(r_{ni} \right) = \frac{\exp\left\{-\alpha d\left(r_{ni}, \rho_{i}\right)\right\}}{\sum_{r_{i} \in [m] \setminus \mathcal{B}_{n}} \exp\left\{-\alpha d\left(r_{i}, \rho_{i}\right)\right\}}.
\end{equation}
\State Set $\mathcal{B}_{n} \gets \mathcal{B}_{n} \cup r_{ni}$.
\State Set $q_{\theta}(\bm{r}_{n} | \mathcal{S}_{n}) \gets q_{\theta}(\bm{r}_{n} | \mathcal{S}_{n}) \times p(r_{ni})$.
\EndFor
\State Define $\bm{r}_{n} \in \mathcal{P}_{m}$ whose $j$th element is $r_{ni}$.
\end{algorithmic}
\end{algorithm}

\subsubsection{Consistent Pairwise Preferences}

When $\bm{y}_{n}$ contains consistent pairwise preferences, $\mathcal{S}_{n}$ is given by \eqref{eq:ConstraintSetPairwise} and contains all topological orderings of the directed acyclic graph given by $\bm{y}_{n}$, or equivalently all linear extensions of the partially ordered set (poset) $\bm{y}_{n}$. \citet{vitelliProbabilisticPreferenceLearning2017} initiated their MCMC algorithm with a single ordering computed deterministically, and then used a modified leap-and-shift algorithm to propose new latent rankings as local perturbations of the current value. This is not sufficient in our case, as we need both the support set of $q_{\theta}(\cdot | \mathcal{S}_{n})$ and its cardinality. 

We will sample latent rankings uniformly on $\text{TO}_{n} = \text{TO}(\bm{y}_{n})$, the set of topological orderings of $\bm{y}_{n}$, and hence need to both count and generate linear extensions. The counting problem itself is known to be \#P complete \citep{brightwellCountingLinearExtensions1991}, although faster algorithms exist for special cases, e.g., sparse posets \citep{kangasCountingLinearExtensions2016}. Generation of the linear extensions can be obtained in constant additional time \citep{pruesseGeneratingLinearExtensions1994}, and very compact storage of the extensions can be obtained using Gray codes \citep{onoConstantTimeGeneration2005,pruesseGeneratingLinearExtensions1994} or permutation decision diagrams \citep{inoueEfficientMethodIndexing2014}. In our implementation we used depth-first search \citep[Ch. 20.3-20.4]{cormenIntroductionAlgorithms2022}, and generated all orderings by looping over all child nodes at each recursive step of the algorithm, keeping track of the solutions via backtracking.

Our procedure for proposing latent rankings from preference data is summarized in Algorithm \ref{alg:LinearExtensions}. Note that for a given $\bm{y}_{n}$ we only need to generate the topological orderings for the items involved in any of the stated pairwise preferences. When all items have been compared ($\bar{\mathcal{A}}_{n} = \emptyset$), the proposal probability is simply one over the number of orderings. When some set of items have not been involved in the comparisons, we consider two settings. First, if all the compared items are preferred to the non-compared items, denoted $\mathcal{A}_{n} \succ \bar{\mathcal{A}}_{n}$ in Algorithm \ref{alg:LinearExtensions}, we permute the non-compared elements and place them after the compared items in the resulting order. This setting is relevant in ranked voting systems. The proposal probability now needs to account for the number of ways of ordering the non-compared items. Finally, if there is no preference relation between the compared and non-compared items, we can insert them in any position in the complete ordering vector, and we need to both account for the number of ways of permuting the uncompared items ($|\bar{\mathcal{A}}_{n}|!$) and the number of ways of inserting them into the complete ordering $\binom{|\mathcal{A}|}{|\bar{\mathcal{A}}_{n}|}$.

\begin{algorithm}[htb]
\caption{Proposing Latent Rankings from Preference Data}
\label{alg:LinearExtensions}
\begin{algorithmic}[1]
\Input{All topological orderings $\text{TO}_{n}$ for items $\mathcal{A}_{n}$, unconsidered items $\bar{\mathcal{A}}_{n} = \mathcal{A} \setminus \mathcal{A}_{n}$.}
\Output{A proposal $\bm{r}_{n}$ and its probability $q_{\theta}(\bm{r}_{n} | \mathcal{S}_{n})$.}
\State Sample an ordering $\bm{o}_{\mathcal{A}_{n}}$ uniformly from $\text{TO}_{n}$.
\If{$\bar{\mathcal{A}}_{n} = \emptyset$}
\State Convert $\bm{o}_{\mathcal{A}_{n}}$ to a ranking $\bm{r}_{n}$ and set $q_{\theta}(\bm{r}_{n} | \theta) = |\text{TO}_{n}|^{-1}$.
\ElsIf{$\mathcal{A}_{n} \succ \bar{\mathcal{A}}_{n}$}
\State Create ordering $\bm{o}_{\bar{\mathcal{A}}_{n}}$ by permuting the items in $\bar{\mathcal{A}}_{n}$ and define $\bm{o}_{n} = (\bm{o}_{\mathcal{A}_{n}}, \bm{o}_{\bar{\mathcal{A}}_{n}})$.
\State Convert $\bm{o}_{\bar{\mathcal{A}}_{n}}$ to ranking $\bm{r}_{n}$ and set $q(\bm{r}_{n} | \theta) = \{|\text{TO}_{n}| \times |\bar{\mathcal{A}}_{n}|!\}^{-1}$.
\Else{}
\State Sample a vector $\bm{\iota}$ of $|\bar{\mathcal{A}}_{n}|$ integers from $\{1,2,\dots,|\mathcal{A}|\}$.
\State Create ordering $\bm{o}_{n}$ with items $\bar{\mathcal{A}}_{n}$ in positions $\bm{\iota}$ and items $\mathcal{A}_{n}$ in the remaining positions.
\State Convert $\bm{o}_{n}$ to ranking $\bm{r}_{n}$ and set $q(\bm{r}_{n} | \theta) = \{|\text{TO}_{n}| \times |\bar{\mathcal{A}}_{n}|! \times \binom{|\mathcal{A}|}{|\bar{\mathcal{A}}_{n}|}\}^{-1}$.
\EndIf
\end{algorithmic}
\end{algorithm}

\section{Generation of Topological Orderings}
\label{sec:TopologicalOrderings}

The generation of all topological orderings when proposing latent rankings in the pairwise preference case is a potential bottleneck. We here report two numerical experiments investigating the extent of this issue in real data. In both experiments we defined $\text{TO}_{n}$ as the number of orderings of the items compared by user $n$, since permuting the non-compared items is a computationally easy task. Computations were performed on a MacBook Pro with a 32 GB Apple M1 Max chip. Further details about the implementation are provided in the first paragraph of Section \ref{sec:SimulationExperiments}.

\subsection{Topological Orderings for PrefLib Data}

We downloaded all datasets containing orders with ties at PrefLib.org \citep{matteiPrefLibLibraryPreferences2013,matteiChapter15APreflib2017}. This included 30 election datasets with the number of votes ranging from 2,477 to 298,788 and the number of candidates between 4 and 23, all donated by \citet{oneillOpenSTV2013}. In addition there was a dataset with 5,000 individuals' ratings of subsets of a total of 100 sushi items \citep{kamishimaNantonacCollaborativeFiltering2003}, and results from an education survey conducted at Instituto Superior Politecnico Jose Antonio Echeverria (Havana, Cuba). 

\begin{figure}[tb]
    \centering
    \includegraphics[width=.75\columnwidth]{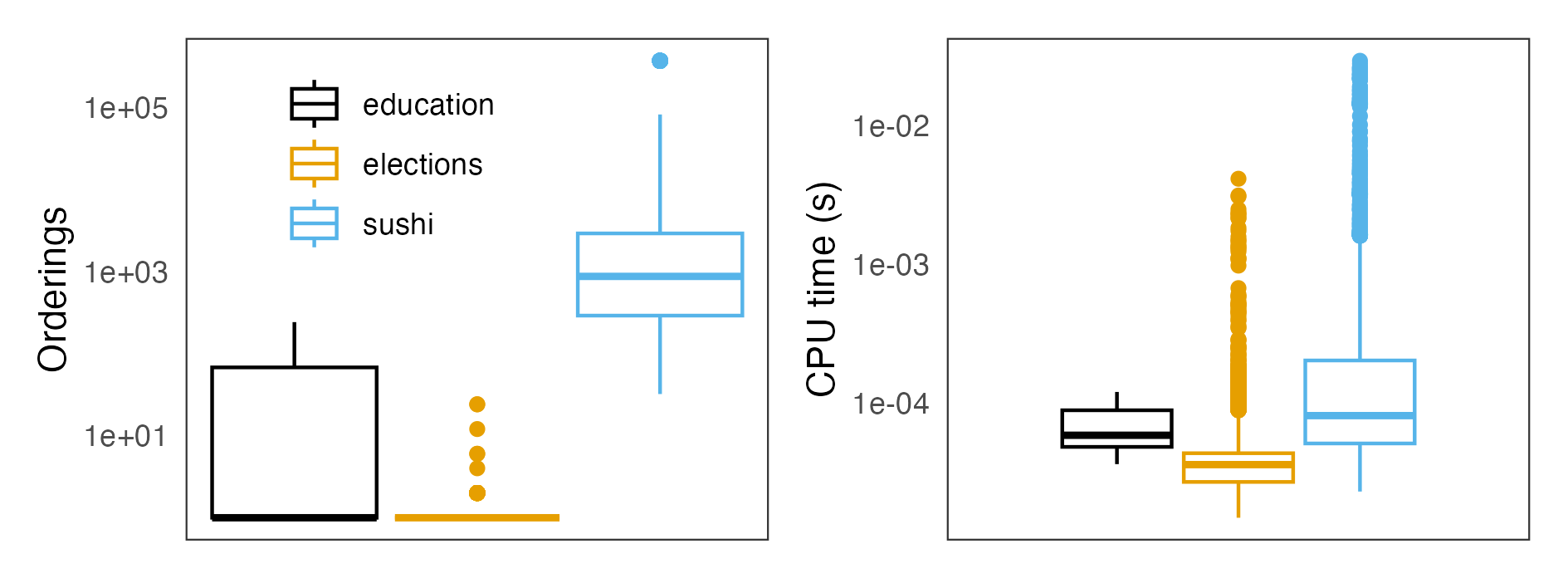}
    \caption{Distributions of the number of topological orderings for each user's preferences  and the required CPU time to compute the orderings.}
    \label{fig:preflib_orderings}
\end{figure}

For all datasets we computed the number of topological orderings $|\text{TO}_{n}|$. The results are shown in Figure \ref{fig:preflib_orderings}, in which the counts for all 30 election datasets have been combined. The largest number of topological orderings occurred for the sushi data, for which the average was $1.2 \times 10^{4}$ and the maximum was $3.6 \times 10^{5}$. The average central processing unit (CPU) time was 0.6 ms, and the maximum was around 30 ms. For the education and election datasets, the average (maximum) CPU times for computing all the orderings of a single user were 0.063 ms (0.12 ms) and 0.041 ms (4.1 ms), respectively. The microbenchmark package \citep{mersmannMicrobenchmarkAccurateTiming2023} was used for the timing.

\subsection{Topological Orderings for Beach Preference Data}

\begin{figure}[tb]
    \centering
    \includegraphics[width=.75\columnwidth]{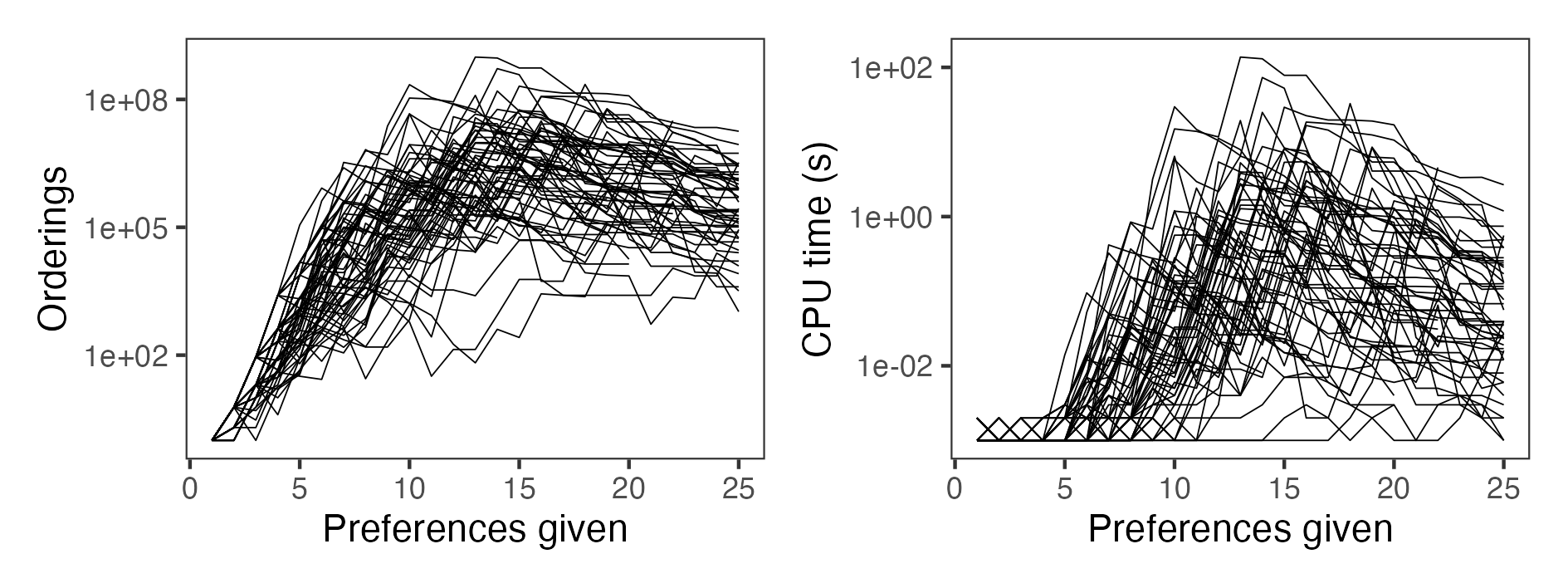}
    \caption{Total number of orderings and associated CPU time as the number of preferences for each user increases, for the beach preference dataset. Each trajectory represents a single user.}
    \label{fig:beach_preference_orderings}
\end{figure}

We studied how the number of orderings depends on the number of stated preferences, using a dataset containing pairwise preferences from 60 users comparing pictures of 15 beaches \citep{vitelliProbabilisticPreferenceLearning2017}. We constructed a temporal order such that each user started with zero preferences, and at each timepoint one new pairwise preference from the user was randomly chosen and added to the user's data. Figure \ref{fig:beach_preference_orderings} shows how the number of orderings and the CPU time developed as more preferences were added. The maximum average CPU time was 4.75 s after 14 preferences, and the maximum CPU time overall was 137 s for a single user after 13 preferences.

\section{Simulation Experiments}
\label{sec:SimulationExperiments}

We here report results of simulation experiments aimed at testing the proposed algorithms. The R packages Rcpp \citep{eddelbuettelSeamlessIntegrationRcpp2013} and RcppArmadillo \citep{eddelbuettelRcppArmadilloAcceleratingHighperformance2014} were used as interfaces to the C++ implementation of our algorithms, which made heavy use of the Armadillo library \citep{sandersonArmadilloTemplatebasedLibrary2016}. Our parallelization of SMC$^{2}$ used the futures framework \citep{bengtssonUnifyingFrameworkParallel2021}, through the furrr package \citep{vaughanFurrrApplyMapping2021}. Pre- and post-processing of data, as well as visualization, was done mainly with the set of R packages provided by the tidyverse \citep{wickhamWelcomeTidyverse2019}.

\subsection{Complete Rankings}
\label{sec:CompleteRankingSimulations}

\begin{figure}[tb]
    \centering
    \includegraphics[width=.9\columnwidth]{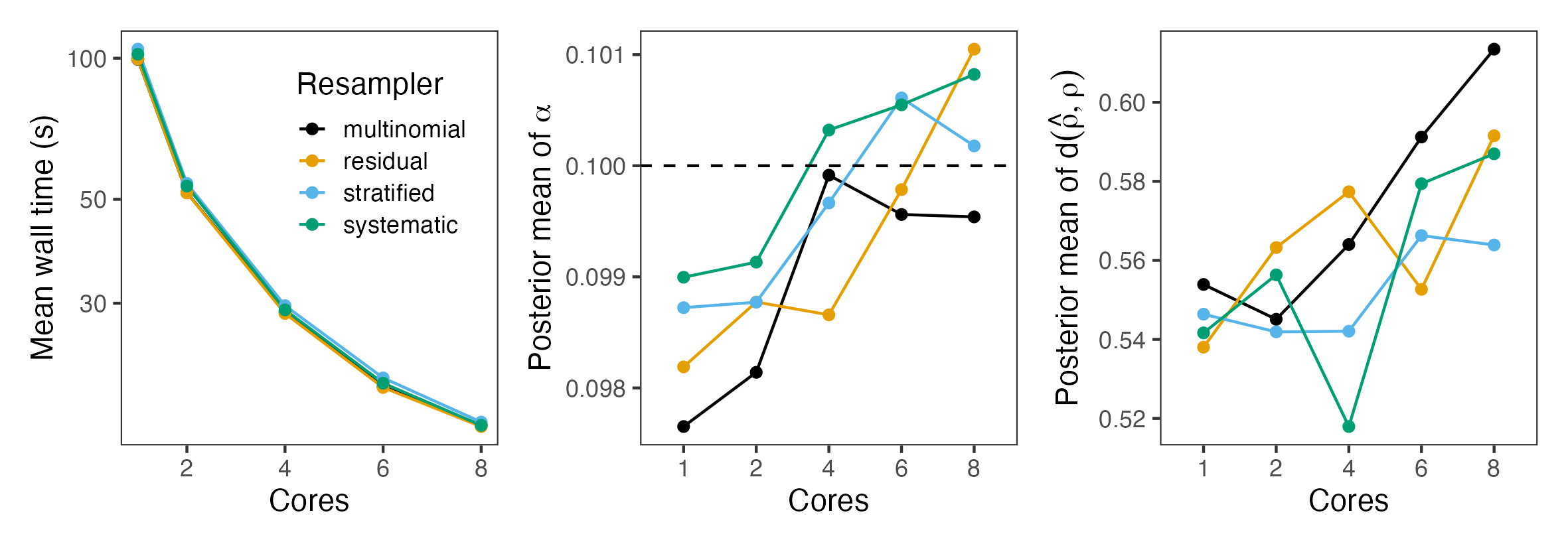}
    \caption{Average wall times (left),  average posterior mean of $\alpha$ (center), and average posterior mean of the footrule distance to the true ranking (right), for simulations with complete rankings.}
    \label{fig:complete-rankings-sequential}
\end{figure}

To study the performance of SMC$^{2}$ in a sequential inference case with complete rankings, we generated 100 datasets with complete rankings from the Mallows model with the footrule distance using the sampling algorithm of \citet[Appendix C]{vitelliProbabilisticPreferenceLearning2017}. In all simulations there were $m=10$ items and $N=1000$ users, the scale parameter was $\alpha=0.1$ and there was a single cluster. The users were assumed to enter one at a time, yielding 1,000 timepoints. For all simulated datasets the algorithm was run in parallel on $P$ cores, with $P\in \{1,2,4,6,8\}$ and $R=5000/P$ particles, using multinomial, residual, stratified, and systematic resampling. The gamma prior for $\alpha$ had shape $\gamma=1$ and rate $\lambda=0.5$ and the resampling threshold was $A = R / 2$. Since the likelihood increments are analytically given by \eqref{eq:CompleteRankingsIncrement}, the number of particle filters was fixed to 1 and the doubling threshold $B$ in Algorithm \ref{alg:SMC2} set to $0$. Simulations were run on the MacBook Pro mentioned in Section \ref{sec:TopologicalOrderings}.

The simulation results are summarized in Figure \ref{fig:complete-rankings-sequential}. The left plot illustrates how the computing time depends on the resampling scheme and parallelization. Regarding the former, all four resampling schemes performed equally fast. Furthermore, the plot shows a clear benefit of parallelization, although with slightly diminishing returns: doubling the number of cores from 1 to 2, 2 to 4, and 4 to 8, respectively, led to factors 1.93, 1.82, and 1.76 speed-up, respectively. The average time with eight cores was 16.5 seconds. For comparison, batch estimation with MCMC with a burnin-in period of 1,000 iterations and 5,000 post-burnin iterations took on average 3.2 seconds. 

The center plot in Figure \ref{fig:complete-rankings-sequential} shows the posterior means of $\alpha$ for different numbers of cores. MCMC batch estimation had average posterior mean at $0.100$ with 95\% Monte Carlo interval $(0.099, 0.101)$, suggesting that the posterior mean on average was very close to the data generating value. SMC$^{2}$ had a slight negative bias\footnote{With bias we mean systematic deviation from the average posterior mean, here computed using MCMC batch estimation with a sufficient number of post burn-in samples. This is not necessarily equal to the data generating value $\alpha=0.1$, although it was very close in this case.} when using a single core, with the upper limit of 95\% Monte Carlo intervals lower than $0.100$. With eight cores there was a tendency towards positive bias, but in this case all the Monte Carlo intervals covered the true value. Finally, the rightmost plot in Figure \ref{fig:complete-rankings-sequential} shows the average posterior mean footrule distance to the true ranking. The differences between cores and resampling methods for the posterior mean of $d(\hat{\bm{\rho}}, \bm{\rho})$ are well within 95\% Monte Carlo confidence intervals, which are not included in the plots for ease of visualization.

\subsection[Top-k Rankings]{Top-$k$ Rankings}
\label{sec:NewUsersTopk}

We next considered the case in which users provide top-3 rankings of $m=10$ items. As in Section \ref{sec:CompleteRankingSimulations}, the users were assumed to arrive sequentially, one at each timepoint. A total of 80 random datasets were simulated, with $N=200$ users, scale parameter $\alpha=0.3$, and footrule distance. Since each user had only ranked three out of ten items, the particle filter in Algorithm \ref{alg:BasicParticleFilter} now had to be run to integrate over the remaining seven items for each user. The total number of particles was set to $R=3000$ which were processed in parallel on 20 cores and combined using \eqref{eq:CombineParallel}. The resampling threshold was $A=R/2$ and the threshold for particle filter doubling was set to an average acceptance rate $B=0.2$ in the rejuvenation step. The initial number of particle filters per core was set to $S=20$. Both uniform and pseudolikelihood proposals were used. Simulations reported in this Section, as well as Sections \ref{sec:PairwisePreferences} and \ref{sec:Mixtures}, were run on the high performance computing cluster Fox provided by the University of Oslo. 

\begin{table}[tb]
    \caption{Results of simulations with sequentially arriving top-3 rankings. Values $\alpha$ and $d(\bm{\rho}, \hat{\bm{\rho}})$ are Monte Carlo averages of the posterior means, with standard errors in parentheses.}
    \label{tab:new_users_topk}
    \begin{tabular}{lllll}
    \toprule
   Proposal & Resampler & $\alpha$ & $d(\bm{\rho}, \hat{\bm{\rho}})$ & Time (minutes) \\    
    \midrule
    \multirow{4}{*}{Pseudolikelihood} & Multinomial & 0.292 (4e-04) & 3.42 (0.023) & 160.2 (0.8)\\
    & Residual & 0.295 (4e-04) & 3.23 (0.021) & 167.1 (0.8)\\
    & Stratified & 0.290 (4e-04) & 3.68 (0.022) & 165.9 (0.8)\\
    & Systematic & 0.295 (4e-04) & 3.70 (0.026) & 158.1 (0.8)\\
    \addlinespace
    \multirow{4}{*}{Uniform} & Multinomial & 0.294 (4e-04) & 3.34 (0.021) & 14.4 (0.1)\\
    & Residual & 0.296 (4e-04) & 3.33 (0.021) & 15.2 (0.1)\\
    & Stratified & 0.296 (5e-04) & 3.40 (0.025) & 15.8 (0.2)\\
    & Systematic & 0.295 (4e-04) & 3.61 (0.027) & 17.2 (0.2)\\
    \bottomrule
    \end{tabular}    
\end{table}

Table \ref{tab:new_users_topk} shows simulation results after the final timepoint. The third column shows that the final estimate of $\alpha$ was close to the true value in all cases. With pseudolikelihood proposal, residual resampling gave the lowest posterior mean distance to the true ranking. With both proposals, systematic resampling gave the highest posterior mean distance to the true ranking. Overall, the performance of uniform and pseudolikelihood proposal seems comparable. Processing a single dataset on eight cores took on average between 14.4 and 17.2 minutes when using uniform proposal, and between 158.1 and 167.1 minutes with pseudolikelihood proposal. In contrast, obtaining 10,000 posterior draws using batch estimation with MCMC took on average five seconds.

\begin{figure}[tb]
    \centering
    \includegraphics[width=.9\columnwidth]{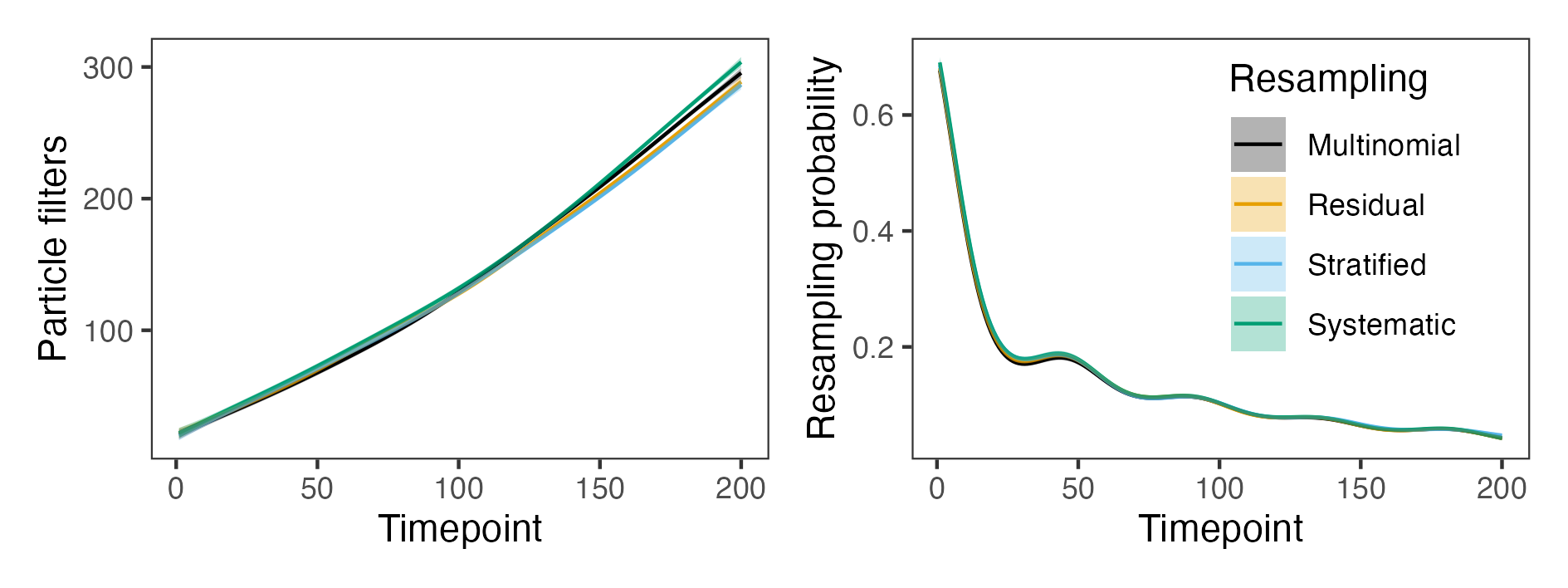}
    \caption{Number of particle filters per core and resampling probability for simulations with sequentially arriving top-3 rankings. Shaded regions are 95\% confidence bands.}
    \label{fig:new_users_topk}
\end{figure}

Figure \ref{fig:new_users_topk} shows the number of particle filters per core as a function of time (left) and the resampling probability per core as a function of time (right). The plots for pseudolikelihood proposal were almost identical, and are not shown. The curves were obtained by fitting generalized additive models (GAMs) \citep{woodGeneralizedAdditiveModels2017} with ten thin-plate regression splines \citep{woodThinPlateRegression2003} as basis functions to the number of particle filters and a binary resampling indicator, respectively. A unit link function was used for the number of particles and a logit link for the resampling indicator, and smoothing was done with restricted maximum likelihood. The left plot shows that the number of particles grows close to linearly. This is expected from Theorem 1 of \citet{andrieuParticleMarkovChain2010}, which implies that the number of particle filters must grow linearly with the number of observations for the acceptance rate to stay constant. The right plot shows that the resampling probability decreases with the number of timepoints. Since each resampling step is followed by one or more rejuvenation steps, this means that as rejuvenation becomes more computationally demanding due to more observations, it also becomes less frequent. This behavior agrees with what is expected from theory; in particular, Theorem 1 and Section 4.3 of \citet{chopinSequentialParticleFilter2002} predicts that the time interval between each time when resampling is needed should increase geometrically in the total number of observations, which would produce an exponentially decaying curve \citep[see also Proposition 17.1 in][]{chopinIntroductionSequentialMonte2020}.

\subsection{Pairwise Preferences}
\label{sec:PairwisePreferences}

\begin{figure}[tb]
    \centering
    \includegraphics[width=\columnwidth]{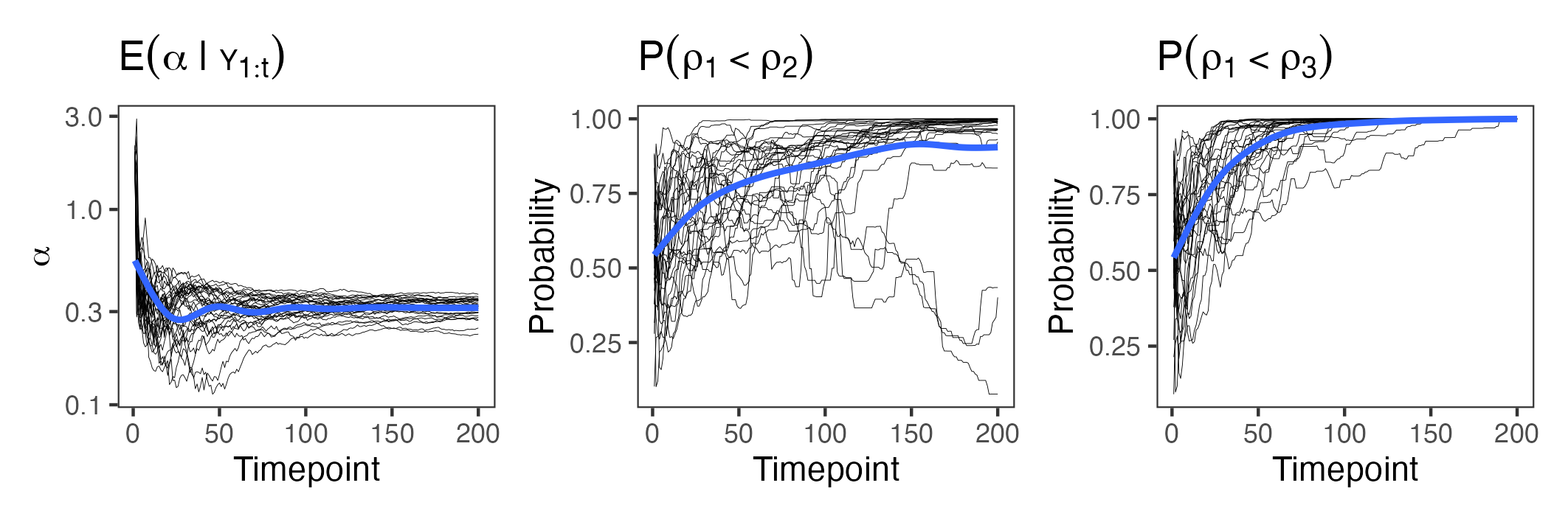}
    \caption{Trace plots of posterior means. Thin black lines shows the posterior probability for a single dataset and thick blue lines are GAM fits.}
    \label{fig:preference-trace}
\end{figure}

We next simulated consistent pairwise preference data by first generating complete rankings of $m=5$ items with $\bm{\rho}=(1,2,\dots,5)'$ and $\alpha=0.3$ for 200 users, and then randomly selecting four implied pairwise preferences for each user. 
The users were assumed to arrive sequentially, one at each timepoint. The process was repeated 30 times, and each dataset was processed in parallel on 30 CPUs, each with $S=100$ particles. The initial number of particle filters was 20, residual resampling was used, and all other parameters were as described in the previous sections. 

Across the 30 datasets, the average posterior mean of $\alpha$ at the final timepoint was $0.329$, with 95\% Monte Carlo interval $(0.327, 0.331)$. For comparison, the average posterior mean of $\alpha$ obtained by MCMC estimation with 20,000 post-burnin iterations on the same datasets was $0.308$ with 95\% Monte Carlo interval $(0.306, 0.310)$, suggesting that SMC$^{2}$ was slightly positively biased in this case. The average wall time for computing the posterior for a simulated dataset was 2.0 minutes, compared to 7.3 seconds for batch estimation with MCMC.

Figure \ref{fig:preference-trace} (left) shows how the posterior expectation of $\alpha$ evolved as more data became available, with initial rapid fluctuations followed by a stabilization close to the true value of $0.3$ after about 100 timepoints. The center plot in Figure \ref{fig:preference-trace} shows the posterior probability that item 1 is preferred to item 2 in the modal ranking, and the right plot show the posterior probability that item 1 is preferred to item 3. As expected, the latter converges more quickly to one than the former. In particular, for three of the simulated datasets $P(\rho_{1} < \rho_{2})$ was below 0.5 even after the final timepoint, implying that items 1 and 2 were "flipped" in the posterior distribution, while $P(\rho_{1} < \rho_{3})$ was close to 1 for all datasets.

\subsection{Mixtures of Mallows Models}
\label{sec:Mixtures}

To test the conditional particle filter and the proposed algorithm's ability to estimate the number of mixture components we randomly generated ten datasets with $m=5$ items and $N=200$ users and ten datasets with $m=5$ items and $N=1000$ users, each having two mixture components with dispersion parameters $\bm{\alpha}=(0.3,0.6)'$ and equal probabilities $\bm{\tau}=(0.5,0.5)'$. Modal rankings were $\bm{\rho}=(1,2,3,4,5)'$ and $\bm{\rho}=(5,4,3,2,1)'$. 

Five models were estimated for each dataset, with $C\in\{1,2,3,4,5\}$ clusters, respectively. For each model and dataset, the SMC$^{2}$ algorithm was run in parallel on 10 CPUs, with $S=400$ particles, each with $R=50$ particle filters. All other parameters were as described for the simulations above. 

\begin{figure}[tb]
    \centering
    \includegraphics[width=.8\columnwidth]{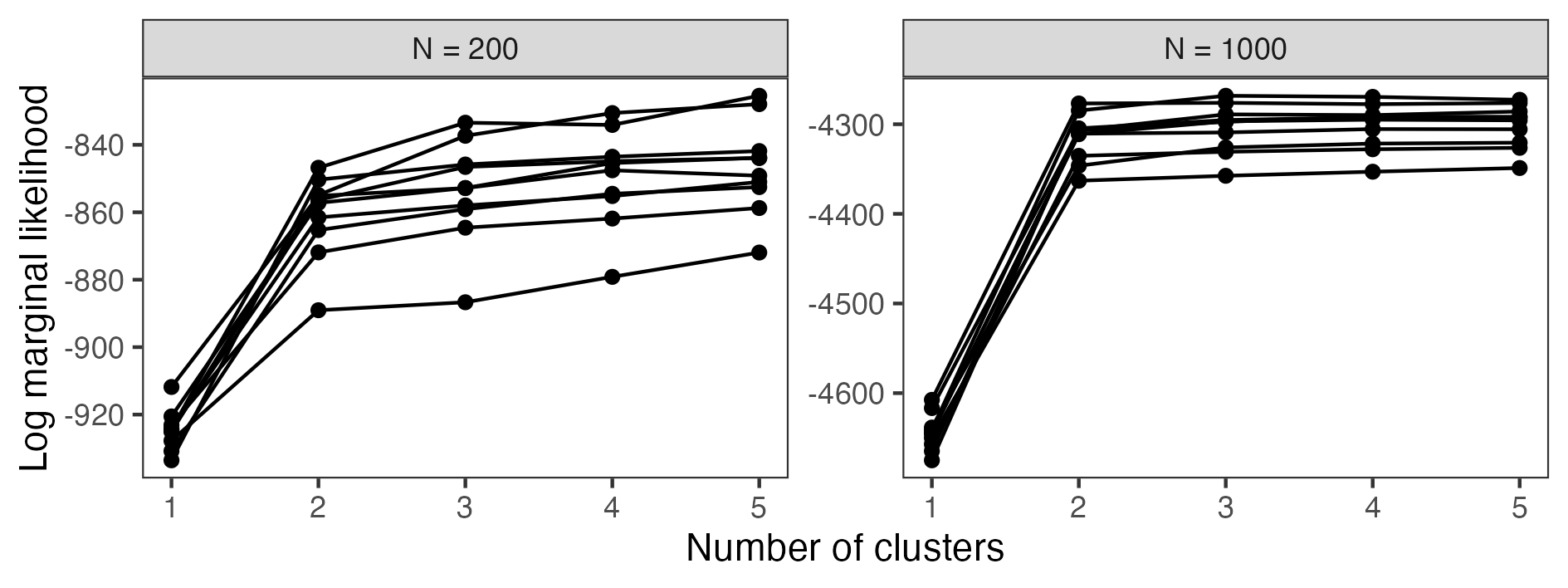}
    \caption{Logarithm of the marginal likelihood as a function of the number of clusters for ten simulated datasets with 200 users (left) and 1000 users (right).}
    \label{fig:clustering-loglik}
\end{figure}

Figure \ref{fig:clustering-loglik} shows how the logarithm of the marginal likelihood \eqref{eq:SMC2MarginalLikelihood} of the models estimated for each dataset varied with the number of clusters. For both the $N=200$ case and the $N=1000$ case we see a clear "elbow" at the correct number of two clusters. In terms of Bayes factors, however, for the $N=200$ case models with more than two clusters were preferred for all the simulated datasets, as can be seen by the fact that the marginal likelihood keeps increasing beyond the two-cluster solution. For the $N=1000$ case, on the other hand, the marginal likelihood is almost completely flat beyond the two-cluster solution. This suggests that this estimator has too high variance in the $N=200$ case. Note that the marginal likelihood is not readily available when estimating mixtures of Mallows models using MCMC algorithms. For example, \citet{vitelliProbabilisticPreferenceLearning2017} and \citet{crispinoBayesianMallowsApproach2019} selected the number of clusters based on finding an "elbow" in a plot of within-cluster distances versus the number of clusters. This method, which requires saving the distance to the cluster centroid for each user at each MCMC step, can also be used as an alternative cluster selection method with SMC.

\begin{figure}[tb]
    \centering
    \includegraphics[width=\columnwidth]{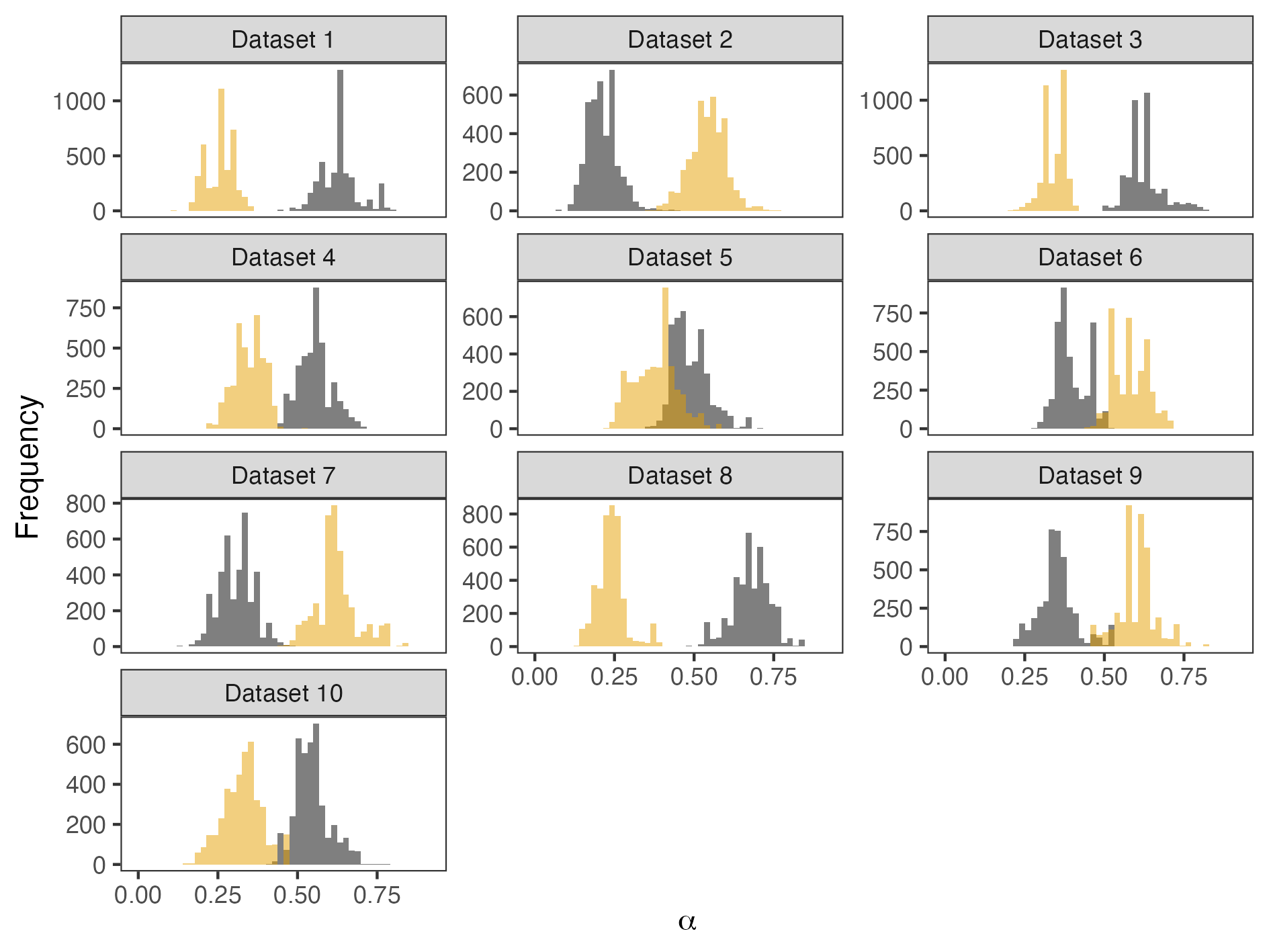}
    \caption{Posterior histograms of $\bm{\alpha}$ for the two-cluster solution, for each simulated dataset with 200 users.}
    \label{fig:clustering-histograms}
\end{figure}

Previous work on estimating mixture models using SMC have either ignored the label switching problem and focused on marginal likelihood estimation \citep{delmoralSequentialMonteCarlo2006,fearnheadParticleFiltersMixture2004} or introduced identifiability constraints \citep{chopinSequentialParticleFilter2002,fearnheadFilteringMethodsMixture2007} despite the known deficits of this approach \citep{jasraMarkovChainMonte2005}. Here we used Stephens' algorithm \citep{stephensDealingLabelSwitching2000} as implemented in the R package label.switching \citep{papastamoulisLabelswitchingPackageDealing2016} for relabeling the outcomes. This has a small additional memory cost, as the cluster probabilities of each user has to be saved at the final timepoint, for each of the $S$ particles. According to the algorithm, there was evidence of label switching in all simulated datasets for all models with $C>1$. Figure \ref{fig:clustering-histograms} shows posterior histograms of the two components of $\bm{\alpha}$ for the two-cluster model. With the exception of dataset 5, the two components were well separated. 

For the two-cluster model with $N=200$, the average of the posterior mean of $\bm{\alpha}$ across the ten simulated datasets was $0.316$ for the smallest component and $0.587$ for the largest component, with probabilities $0.502$ and $0.498$, both very close to the values in the data generating distribution. The average posterior probability of the true modal ranking $\bm{\rho}_{c}$ in each mixture component was $0.855$ in the cluster with $\alpha_{c}=0.3$ and $1.000$ in the cluster with $\alpha_{c}=0.6$, confirming that the modal ranking is easier to identify with a higher precision parameter. With $N=1000$, the average posterior mean of $\bm{\alpha}$ was $(0.303, 0.612)'$, with probabilities $(0.500, 0.500)'$. The average posterior probability of the true modal ranking was $1.000$ for each component.

\subsection{Timing Comparisons for Sequential Estimation}

To compare the relative speed of SMC$^{2}$ and MCMC for sequential estimation, we performed an additional set of experiments in which a set of users were assumed to arrive one at a time, and the posterior had to be updated after the arrival of each new user. For MCMC, this requires rerunning the whole algorithm at each timepoint, whereas for SMC$^{2}$ a single iteration of Algorithm \ref{alg:SMC2} is sufficient. 

For a total number of timepoints equal to 10, 30, 50, 70, and 90 we generated 100 random datasets with $m=10$ items and $\alpha=0.1$. In addition to the complete rankings case, we also did experiments in which only the top-5 or the top-3 items were retained. For each simulated dataset, 5,000 samples from the posterior distribution were obtained. For MCMC we used a burn-in of 1000 and no thinning, so that a total of 6,000 iterations were run each time a new observation arrives. For SMC$^{2}$, the number of particles was set to 5,000 which were either processed on a single core or in parallel on eight cores. The initial number of particle filters was set to 1, and the thresholds for resampling and particle filter doubling were set identically to Section \ref{sec:NewUsersTopk}. We emphasize that it is hard to ensure that the effective posterior sample size obtained from SMC$^{2}$ and MCMC are identical and hence obtain decisive evidence of which algorithm is faster. Our goal here was rather to understand the scaling behavior of the respective algorithms for sequential estimation, and how this depends on the amount of missing data.

\begin{figure}[tb]
    \centering
    \includegraphics[width=\columnwidth]{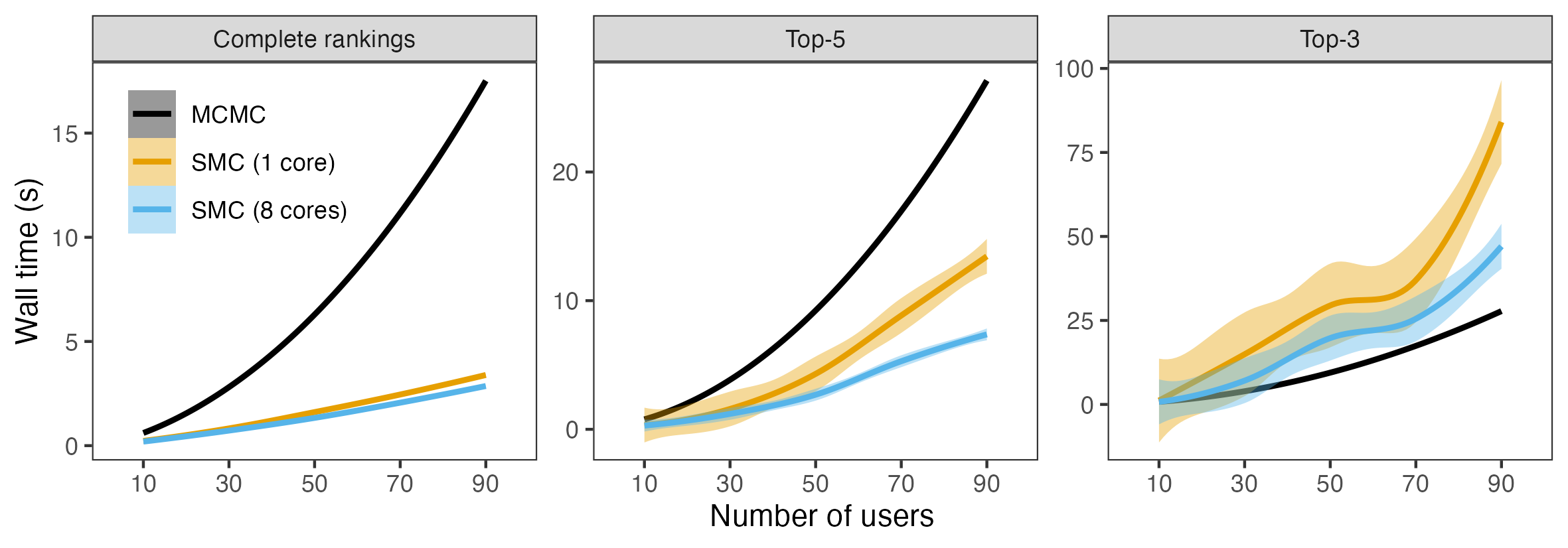}
    \caption{Wall time for sequential estimation with complete rankings (left), top-5 rankings (center), and top-3 rankings (right). Shaded areas are 95\% confidence intervals.}
    \label{fig:timing-comparisons}
\end{figure}

The results are shown in Figure \ref{fig:timing-comparisons}. In all cases, MCMC scaled approximately quadratically. On the other hand, SMC on a single core and on eight cores scaled close to linearly with complete rankings. In the case with missing rankings, SMC scaled quadratically, but remained faster than MCMC in the top-5 case. In the top-3 case, however, the computing time for SMC$^{2}$ grew more quickly with the number of users than MCMC. Closer examination of the data revealed that this was due to the particle filter doubling required to properly integrate over the latent variables in these cases.

\section{Sequential Analysis of Formula 1 Data}
\label{sec:applications}

We now analyze the Formula 1 data introduced in Figure \ref{fig:f1-data-plot}. As mentioned in the introduction, the rankings were derived from the results of each race. Whenever a driver either was disqualified, did not finish a race, or did not start, the ranking for the particular race was set to missing and assumed to have a higher value than all the ranked drivers. This yielded a missingness proportion of 12.7\%. Each race $n \in \{1, \dots, 68\}$ was hence modeled as an assessor yielding a top-$k_{n}$ ranking, with $k_{n}$ equal to the number of drivers completing the race.

We computed the posterior distribution of the Bayesian Mallows model with footrule distance sequentially using the SMC$^{2}$ algorithm with $R=10^{5}$ particles and $S=10$ initial particle filters. We used multinomial resampling, uniform latent rank proposal, resampling threshold $A = R/2$, and threshold $B=0.2$ for particle filter doubling. We confirmed that the number of particle filters was sufficiently large by running the model multiple times with different random number seeds and checking that the posterior quantities of interest remained essentially the same. At the final timepoint, the posterior mean of the scale parameter $\alpha$ was 0.170 with 95\% posterior interval $(0.165, 0.178)$. 

\begin{figure}
    \centering
    \includegraphics[width=\linewidth]{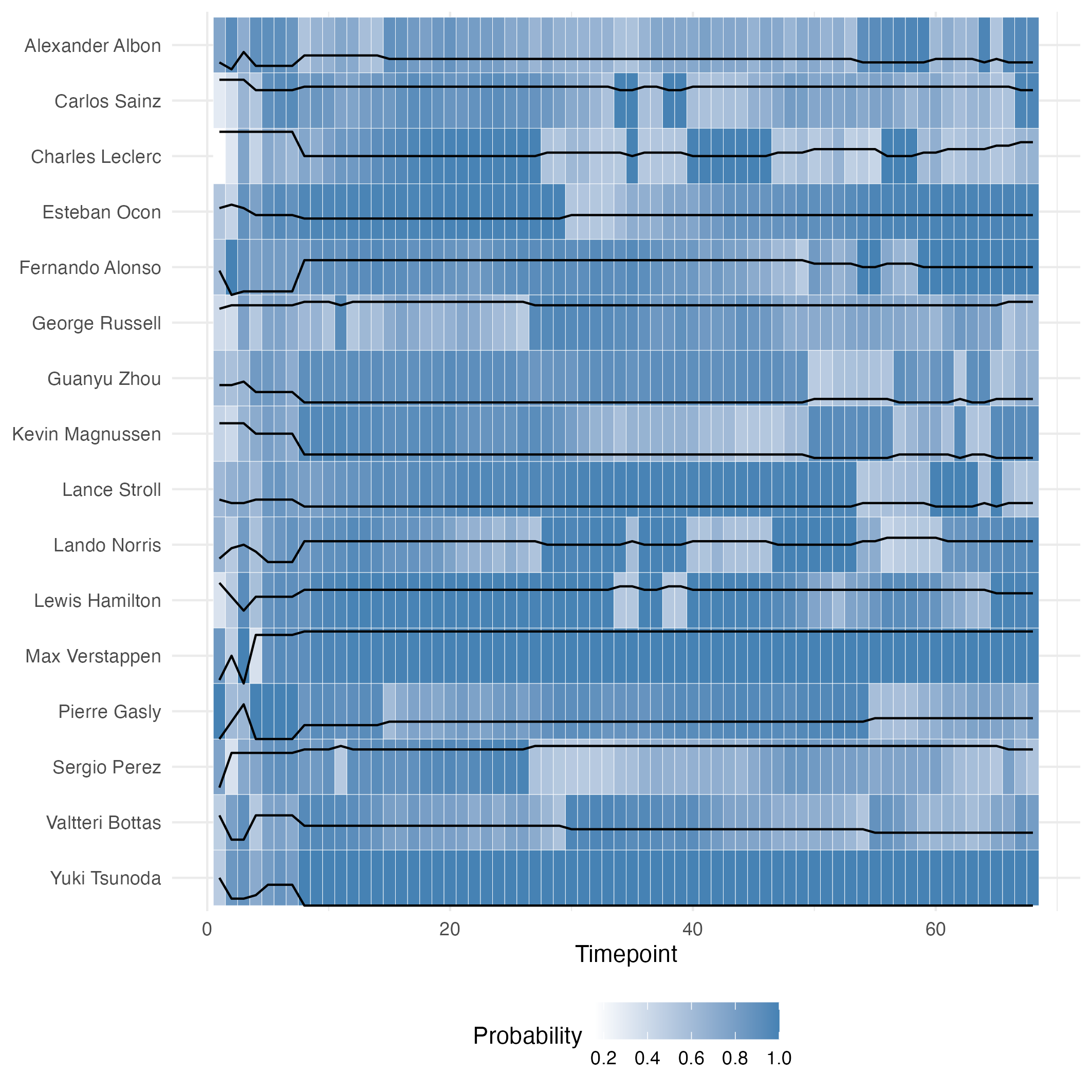}
    \caption{CP consensus over time for Formula 1 drivers. The black lines show the CP consensus ranking, and the color scale shows the posterior probability of having the given ranking or higher.}
    \label{fig:f1-cp-consensus}
\end{figure}

At each timepoint we computed the cumulative probability (CP) consensus based on the current posterior distribution. The CP consensus \citep[Sec. 5.1]{vitelliProbabilisticPreferenceLearning2017} was computed by first selecting the driver with the highest posterior probability of being ranked first. Then among the remaining drivers we found the driver with the highest posterior probability of being ranked first or second, and so on. The resulting trace plots are shown in Figure \ref{fig:f1-cp-consensus}, in which the solid black lines indicate the CP consensus ranking of each driver and the color scale indicates the posterior probability. Note that by construction, the posterior probability for the driver whose CP consensus equals 16 -- the number of items -- always has probability 1, since all drivers must be ranked 16 or higher. Also note that the true ranking $\bm{\rho}$ is assumed to be constant, and that Figure \ref{fig:f1-cp-consensus} shows how our posterior belief about this parameter develops as we obtain more data.

Some interesting features can be seen from Figure \ref{fig:f1-cp-consensus}. First we note that the CP consensus changes quickly during the first ten races, as more data is incorporated into the posterior and the effect of the uniform prior fades. Correspondingly, the probability for the drivers ranked as best after the first rounds is relatively low. For example Charles Leclerc is ranked first in the first eight timepoints, with a probability between 0.16 and 0.76 and Carlos Sainz is ranked second during the first three timepoints, with a probability between 0.29 and 0.65. After the first ten races the consensus stabilizes. For example, Max Verstappen is ranked first for the first time after the eighth race with a probability of 0.91 and retains this rank until the last timepoint with a probability exceeding 0.99 from the thirteenth timepoint and onward. However, there is also movement in the later timepoints; for example, Charles Leclerc climbs from rank 8 at timepoint 54 to rank 4 at the final timepoint.

\section{Discussion}
\label{sec:Discussion}

We have proposed an SMC$^{2}$ algorithm for sequential estimation of the Bayesian Mallows model. The algorithm naturally incorporates data in the form of partial rankings and both transitive and non-transitive pairwise preferences, and is straightforward to parallelize. Compared to MCMC, the algorithm is competitive in use cases where data arrive sequentially and the posteriors of interest need to be recomputed for each new data batch. In batch estimation problems, on the other hand, MCMC has been faster in all cases considered in this paper. 

A number of future extensions are possible. First, conditioning on the full particle history in the conditional particle filter may lead to a high degree of degeneracy \citep{Whiteley:2010}, with the consequence that a large number of particle filters is required to obtain a sufficiently accurate approximation of the conditional posterior. Backward sampling has been shown to considerably decrease this degeneracy \citep{lindstenUseBackwardSimulation2012}, and would be interesting to consider in the current setting. Next, a challenge with SMC$^{2}$ is that its computational complexity grows quadratically with time, as the plot for top-3 rankings in Figure \ref{fig:timing-comparisons} (right) illustrates. A related nested particle filter algorithm which scales linearly with time has been proposed by \citet{crisan_nested_2018}. Since their algorithm requires the parameters to be real-valued, it is not directly applicable to the Mallows model, but creating such an extension is an interesting problem for future research. Another extension of practical interest is to allow users to provide updated rankings, e.g., comparisons of previously unseen items. \citet[Ch. 6]{steinSequentialInferenceMallows2023} proposed an algorithm for this in the case of partial rankings, in which users whose new data contradicted their current latent rankings were removed from the pool and then reentered. An extension of this approach to SMC$^{2}$ would likely require a particle filter which performs this type of correction while still yielding unbiased estimates of the marginal likelihood $p(\bm{y}_{\mathcal{I}_{1:t}} | \theta)$, as equation \eqref{eq:Zhat} does in the current case where new users arrive at each timepoint. Finally, it may be of interest to let some or all of the parameters in $\theta$ depend on time, e.g., to monitor how preferences in a population evolve. An MCMC algorithm for time-varying modal ranking $\bm{\rho}$ has been proposed by \citet{asfawTimevaryingRankingsBayesian2017}, but sequential estimation is likely a good alternative in this case. More recently \citet{piancastelli_time_2025} have proposed a model for timeseries of rankings, based on the Mallows model. In this model, complete or partial rankings of a set of items are assumed to be observed on a relatively large number of timepoints, and rather than estimating the consensus ranking the focus is on timeseries parameters describing the dynamics by which the modal ranking changes over time. 

Another interesting possibility is to use the proposed algorithm in a sequential experimental design framework. For example, at a given timepoint, the items to be ranked or compared by the next user could be determined by a utility function seeking to maximize the information about some posterior quantity of interest, e.g., whether an item $A_{1}$ is preferred to another item $A_{2}$ in the modal ranking. Examples of similar uses of SMC include estimation of generalized (non-)linear models \citep{drovandiSequentialMonteCarlo2013}, model selection \citep{drovandiSequentialMonteCarlo2014}, and hierarchical models \citep{mcgreePseudomarginalSequentialMonte2016}.

\section*{Acknowledgement}

The authors thank Arnoldo Frigessi for discussions and encouragement. Ø.S. thanks Marta Crispino for fruitful discussions about rank modeling.

\bibliographystyle{apalike}
\bibliography{references}

\end{document}